\documentclass[twocolumn,showpacs,preprintnumbers,amsmath,amssymb]{revtex4}

\usepackage{graphicx}
\usepackage{dcolumn}
\usepackage{bm}

\begin{document}


\title {Chaotic motion of neutral and charged particles in the magnetized Ernst-Schwarzschild spacetime} 
\author{Dan Li}
\author{Xin Wu}
\email{xwu@ncu.edu.cn}
\affiliation{Department of Physics $\&$ Institute of Astronomy,
Nanchang University, Nanchang 330031, China}


\begin{abstract}

Neutral test particles around a Schwarzschild black hole immersed in an external uniform magnetic
field have no interactions of electromagnetic forces,
but their motions can be chaotic. This chaotic behavior is induced owing to the gravitational effect of the magnetic field
leading to the nonintegrability of the magnetized Ernst-Schwarzschild spacetime geometry.
In fact, chaos is strengthened typically with an increase of
the energy or the magnetic field under appropriate circumstances.
When these test particles have charges, the electromagnetic forces are included.
As a result, the electromagnetic forces have an effect on strengthening or weakening
the extent of chaos caused by the gravitational effect of the magnetic field.

\end{abstract}


\maketitle

\section{Introduction}

Schwarzschild, Kerr and Kerr-Newman black hole spacetimes
are highly nonlinear, but they have a sufficient number of isolating integrals to separate equations of motion and therefore are
strictly integrable, regular. Only when perturbations are included and destroy
the integrability of these spacetime geometries, may chaos  be induced. Here the `chaos' means that
the dynamics of test particles in these spacetimes display exponentially sensitive dependence on initial conditions.
In addition, the perturbations come from materials (such as disks, massive halos, shells and rings) surrounding
the central gravitational bodies, spins of the particles or the central bodies, and so on.
In fact, chaos can occur in static, axially symmetric relativistic core-shell systems with a monopolar core and
an exterior shell of dipoles, quadrupoles and octopoles [1-3].
The chaotic motions of particles around black holes with discs or rings were also confirmed in [4-6].
Systems of compact binaries with one or two bodies spinning can be chaotic in some cases.
For reference, see e.g. [7-13].

In addition to the above-mentioned materials and spins, magnetic or electromagnetic fields
act as an additional perturbation to the spacetime geometries for inducing chaos.
The magnetized Ernst-Schwarzschild metric [14] is a static, axially symmetric  spacetime,
and describes the motion of electrically charged and
neutral test particles around a Schwarzschild black hole immersed in an external uniform magnetic
field. Employing the methods of Poincar\'{e} surfaces of section and Lyapunov exponents, Karas $\&$ Vokroulflick\'{y} [15]
showed that higher energy yields stronger chaotic behavior
regardless of whether the particles in this spacetime geometry have charges or not.
In fact, this chaotic behavior can be induced because no additional constant of motion exists
and the equations of motion of test particles cannot be successfully separated and solved with an analytical method.
Recently, Wang et al. [16] investigated the chaotic motion of  a
test scalar particle coupling to the Einstein tensor in Schwarzschild-Ernst black hole spacetime.
When an external magnetic field whose four-vector potential
have two nonzero components $A_{t}$ and $A_{\phi}$ is included near the Kerr black hole, the motion of charged particles in
this  axially symmetric  spacetime geometry is nonintegrable
and probably chaotic [17-19].
If the four-vector potential of a magnetic field has  four nonzero components and breaks axial symmetry,
the transition from regular to chaotic dynamics occurs more easily [20].

As claimed in [15], the gravitational effect of the magnetic
field can induce chaos although neutral particles in the magnetized Ernst-Schwarzschild spacetime
have no interactions of electromagnetic forces. It was also found in [15] that
higher energy leads to stronger chaotic behavior.
However, one question remains how the extent of chaos changes with an increase of the magnetic
field. Another question remains whether the chaotic behavior caused by the gravity of the magnetic
field is strengthened when electrically charged particles have interactions of electromagnetic forces.
To address these questions,
we employ a geometric numerical
integration algorithm with preservation of geometric properties of the flows of the differential
equations of motion of particles. Manifold correction schemes [21-25],
symplectic integrators [26-29]
and symmetric methods [30] are some of the geometric algorithms. Because the equations of motion are inseparable in
the present problem, one of the explicit symmetric algorithms in extended phase space [31-35] is considered as a numerical integration tool.
On the other hand,  methods to determine the onset of chaos should be
unambiguous declarations of chaos. That is, they do
not depend on the  spacetime coordinate system used in general relativity. They are called invariant chaotic indicators.
The methods of poincar\'{e} sections,
Lyapunov exponents of two nearby orbits with the proper time and distances [36,37],
and fast Lyapunove indicators of two nearby orbits with the proper time and distances [37]
are what we expect to use. In a word, the fundamental aim of this paper is to apply an appropriate geometric integrator and
unambiguous indicators of chaos to truly describe the effect of the electromagnetic forces on chaos induced by the gravity of the magnetic
field.

In what follows, we introduce the magnetized Ernst-Schwarzschild spacetime and discuss
the effective potential and stable circular orbits at the equatorial plane in Section 2.
Then, we survey the dynamics of generic orbits using analytical or numerical methods in Section 3.
Finally, the main results of this paper are concluded in Section 4.

\section{Magnetized Ernst-Schwarzschild spacetime}

In this section, we introduce the evolution model of charged particles in the magnetized Ernst-Schwarzschild spacetime geometry
and an electromagnetic field. Then, the effective potential and innermost stable circular orbits at the equatorial plane are discussed.

\subsection{Dynamical model of charged particles}

Ernst [14] described  the motion of neutral particles around a non-rotating black hole in an external magnetic field
using the Ernst-Schwarzschild metric
\begin{eqnarray}
ds^{2} &=& g_{\alpha\beta}dx^{\alpha}dx^{\beta} \nonumber \\ &=&
\Lambda^{2}[\frac{2M-r}{r}dt^{2}+
\frac{rdr^{2}}{r-2M}
 +r^{2}d\theta^{2}] \nonumber \\
 &&  +\frac{r^{2}}{\Lambda^2}\sin^{2}{\theta}d\phi^{2},
\end{eqnarray}
where $\Lambda  = 1+(1/4)B^{2}r^{2}\sin^{2}\theta$, $B$ is a magnetic field parameter, and $M$
stands for the mass of the black hole. The superscripts of $x$, $x^{\alpha}$ and $x^{\beta}$ with $\alpha,\beta \in \{0,1,2,3\}$,
represent the components
of Schwarzschild-like coordinates $\mathbf{x}=(x^{0},x^{1},x^{2},x^{3})=(t,r,\theta,\phi)$.
This metric is an invariant line element in the
four-dimensional spacetime. In fact, $ds^{2}=-d\tau^{2}$, where $\tau$ stands for the  proper time.
Here the constant of gravity $G$ and the speed of light $c$ use the geometric units $c=G=1$.
The spacetime has an event horizon at $r=2M$ and no singularity outside the horizon.
It is the same as the Schwarzschild  metric in the two points.
However, the difference between them lies in that the Schwarzschild spacetime is
asymptotically flat, but this Ernst-Schwarzschild spacetime is not due to
the magnetic field acting as a gravitational effect.
When the variables and parameters are given to
scale transformations
$t\rightarrow tM$, $r\rightarrow rM$, $B\rightarrow B/M$ and $\tau\rightarrow \tau M$,
we obtain a dimensionless form of the metric, which corresponds to the dimensionless  Lagrangian
\begin{eqnarray}
\mathcal{L} &=& \frac{1}{2}(\frac{dS}{d\tau})^{2}=\frac{1}{2}g_{\alpha\beta}\dot{x}^{\alpha}\dot{x}^{\beta} \nonumber \\
 &=& \frac{1}{2}\Lambda^{2}[-(1-\frac{2}{r})\dot{t}^{2}+(1-\frac{2}{r})^{-1}\dot{r}^{2}+r^{2}\dot{\theta}^{2}] \nonumber \\
 & &     +\frac{1}{2}\Lambda^{-2}r^{2} \sin^{2}{\theta}\dot{\phi}^2.
\end{eqnarray}
Notice that $\mathcal{L}$ is identical to -1/2, $\mathcal{L}=-1/2$. We define
the covariant momentum
\begin{equation}
P_{\alpha}=\frac{\partial\mathcal{L}}{\partial\dot{x}^{\alpha}}=g_{\alpha\beta}\dot{x}^{\beta}.
\end{equation}
This Lagrangian is exactly equivalent to the  Hamiltonian
\begin{equation}
\mathcal{H} =\frac{1}{2}g^{\alpha\beta}P_{\alpha}P_{\beta}.
\end{equation}
However, Lagrangian and Hamiltonian formulations at the same post-Newtonian order
are not exactly equivalent and are approximately related in general [38-40].
The difference between them is mainly due to
higher-order  post-Newtonian terms truncated.

On the other hand,
Ernst [14] provided the non-zero components of the magnetic field
\begin{eqnarray}
B_{r} &=& \Lambda^{-2}B\cos\theta, \nonumber \\
B_{\theta} &=& -\Lambda^{-2}B(1-\frac{2}{r})^{1/2}\sin\theta.
\end{eqnarray}
The electromagnetic tensor $F_{\mu\nu}$ contains non-vanishing components [41]
\begin{eqnarray}
F_{\theta\phi} &=& \frac{1}{2}B\Lambda^{-2}r^{2}\sin{2\theta},  \nonumber \\
F_{r\phi} &=& B \Lambda^{-2}r\sin^{2}{\theta}.
\end{eqnarray}
Because the four-vector potential \textbf{A} and the electromagnetic tensor $F_{\mu\nu}$
satisfy the relation $F_{\mu\nu}=A_{\nu,\mu}-A_{\mu,\nu}$,
we have the four-vector potential with a non-zero component
\begin{equation}
A_{\phi} = \frac{B}{2\Lambda}r^{2}\sin^{2}{\theta}.
\end{equation}
The potential used in this paper is the same as that used in [42] but
is slightly different from those of [15] and [16]. That is,
$A_{\phi} =- Br^{2}\sin^{2}{\theta}/(2\Lambda)$ in [15]
and $A_{\phi} =- Br^{2}\sin^{2}{\theta}/\Lambda$ in [16].
Let $q$ be an electric charge of the test particle with the dimensionless operation $q\rightarrow qM$.
The charged  particle moves under  the gravities of the system (2) and  the electromagnetic force,
and its covariant momentum obeys the relation
\begin{equation}
p_{\alpha}=P_{\alpha}+qA_{\alpha}.
\end{equation}
Based on Eqs. (4) and (8), the motion of the charged
particle  subject to the influences of the gravities and  the electromagnetic force
is restrained by the Hamiltonian
\begin{equation}
\mathbb{H} =\frac{1}{2}g^{\alpha\beta}(p_{\alpha}-qA_{\alpha})(p_{\beta}-qA_{\beta}).
\end{equation}
Hamilton's canonical equations of motion are given as
\begin{equation}
\frac{dx^{\alpha}}{d\tau} =\frac{\partial \mathbb{H}}{\partial p_{\alpha}}, ~~~~
\frac{dp_{\alpha}}{d\tau} =-\frac{\partial \mathbb{H}}{\partial x^{\alpha}}.
\end{equation}
Both electromagnetic field and spacetime geometry are stationary and axial symmetric,
i.e., the Hamiltonian $\mathbb{H}$ does not explicitly depend on $t$ and $\phi$.
Therefore, Eq. (10) implies that the system $\mathbb{H}$
has constant  specific  energy $E$ and angular momentum $L$ as follows:
\begin{eqnarray}
p_{t} &=& P_{t}+qA_{t}=P_{t}=g_{tt}\dot{t} \nonumber \\
&=& -\Lambda^{2}(1-\frac{2}{r})\dot{t}=-E, \nonumber \\
p_{\phi} &=& P_{\phi}+qA_{\phi}=g_{\phi\phi}\dot{\phi}
+qA_{\phi} \nonumber \\
&=& \Lambda^{-2}r^{2} \sin^{2}{\theta}\dot{\phi}+\frac{qBr^{2}\sin^{2}{\theta}}{2\Lambda} \nonumber \\
&=& L.
\end{eqnarray}
This constant angular momentum $L$ exists because the system (9)  preserves axial symmetry.
However, there is no constant angular momentum in the system of [20] for the lack of axial symmetry.
In terms of the two constants of motion, the Hamiltonian $\mathbb{H}$ is rewritten as
\begin{eqnarray}
H &=& \frac{1}{2}[\frac{r-2}{r\Lambda^{2}} p^{2}_{r}+\frac{p^{2}_{\theta}}{r^{2}\Lambda^{2}}
+\frac{rE^{2}}{(2-r)\Lambda^{2}} \nonumber \\
&& +\frac{\Lambda^{2}}{r^{2} \sin^{2}\theta} (L-\frac{qBr^{2}\sin^{2}{\theta}}{2\Lambda})^{2}].
\end{eqnarray}
The system $H$ has a four-dimensional phase space made of $(r,\theta, p_{r}, p_{\theta})$.
Its canonical equations of motion are
\begin{eqnarray}
& & \frac{dr}{d\tau} =\frac{\partial H}{\partial p_{r}}, ~~~~
\frac{dp_{r}}{d\tau} =-\frac{\partial H}{\partial r}, \nonumber \\
& & \frac{d\theta}{d\tau} =\frac{\partial H}{\partial p_{\theta}}, ~~~~
\frac{dp_{\theta}}{d\tau} =-\frac{\partial H}{\partial \theta}.
\end{eqnarray}
Obviously, the Hamiltonian itself is an integral of motion. In fact, it is always identical to -1/2,
\begin{equation}
H=-\frac{1}{2}.
\end{equation}

\subsection{Effective potential and innermost stable circular orbits}

Noting Eqs. (12) and (14), we have
\begin{eqnarray}
(1-\frac{2}{r})^2p_{r}^2+r^{-2}(1-\frac{2}{r})p_{\theta}^2=E^2-V^2,   \nonumber \\
V^2=(1-\frac{2}{r})[1+\frac{\Lambda^2}{r^2}(L-\frac{qBr^2}{2\Lambda})^2]\Lambda^2,
\end{eqnarray}
where $V$ is the effective potential  at the equatorial plane $\theta$=$\pi/2$.
Fig. 1 plots the effective potentials for various values of the parameters $B$, $q$ and $L$.
When $V$ has its local minimum, $\dot{V}=0$, that is, a circular orbit exists.
In this case, the value of $r$ is the radius of the circular orbit.
If  $\ddot{V}>0$, then this circular orbit is stable. Table 1 lists the radii of the
stable circular orbits in Fig. 1.
Especially for $\ddot{V}=0$, the stable circular orbit obtained is the innermost stable circular orbit (ISCO).
For example, $r=6$ is the radius of ISCO of the system (12) with $B=0$, equivalently, that of
the Schwarzschild spacetime. Since the magnetic or electromagnetic field we consider is weak, i.e., $0<|B|\ll 1$
and $0<|qB|\ll 1$, the radii of ISCOs approach to 6 for the various values of the parameters $B$ and $q$, as shown in Fig. 2 and Table 2.
In the following discussions, we apply analytical or numerical methods to study the dynamics of generic orbits.

\section{Investigations of orbital dynamics}

We explore the dynamics of the system (12) according to three cases: $B=0$, $B\neq0$ with $q=0$, and $B\neq0$ with $q\neq0$.

\subsection{Case 1: $B=0$}

For the case of $B=0$, the system (12) corresponds to the Schwarzschild spacetime
\begin{eqnarray}
H = \frac{1}{2}(\frac{r-2}{r} p^{2}_{r}+\frac{p^{2}_{\theta}}{r^{2}}
+\frac{rE^{2}}{2-r}+\frac{L^{2}}{r^{2}\sin^{2}\theta}).
\end{eqnarray}
Noting Eq. (14) and the Hamilton-Jacobi equation, we have
\begin{eqnarray}
-1 &=& \frac{r-2}{r} (\frac{\partial S}{\partial r})^{2}+\frac{1}{r^{2}}(\frac{\partial S}{\partial \theta})^{2}
+\frac{rE^{2}}{2-r} \nonumber \\
&& +\frac{L^{2}}{r^{2}\sin^{2}\theta},
\end{eqnarray}
where $S$ is a generating function of the form
\begin{equation}
S=\frac{1}{2}\tau+L\phi+S_{1}(r)+S_{2}(\theta).
\end{equation}
It is clear that Eq. (17) has a separable form of the variables $r$ and $\theta$
and can be split into two equations
\begin{eqnarray}
K &=& r^{2}-(1-\frac{2}{r})^{-1}E^2r^2+(1-\frac{2}{r})r^2(\frac{\partial S_{1}}{\partial r})^2, \nonumber\\
K &=& -(\frac{\partial S_{2}}{\partial \theta})^2-\frac{L^2}{\sin^2\theta},
\end{eqnarray}
where $K$ is a constant.
Considering that $\partial S_{1}/\partial r=p_r=r\dot{r}/(r-2)$ and $\partial S_{2}/\partial \theta=p_{\theta}=r^{2}\dot{\theta}$,
we modify Eq. (19) as
\begin{eqnarray}
K &=& r^{2}+r^{3}(\dot{r}^{2}-E^{2})/(r-2), \\
K &=& -(r^{2} \dot{\theta})^2-\frac{L^2}{\sin^2\theta}.
\end{eqnarray}
Obviously, $(r,\theta,\dot{r},\dot{\theta})$ is easily, analytically solved.
In other words, the 4-dimensional Hamiltonian (16) (i.e. the Schwarzschild spacetime) is integrable due to the existence
of the two constants of motion given by Eqs. (14) and (20) [or (21)].
All non-circular  orbits of test particles in this spacetime are quasi-periodic and regular.
Of course,  all circular orbits should be strictly periodic.

\subsection{Case 2: $B\neq0$, $q=0$}

When $B\neq0$ and $q=0$, Eq. (12) becomes
\begin{eqnarray}
H &=& \frac{1}{2\Lambda^{2}}(\frac{r-2}{r} p^{2}_{r}+\frac{p^{2}_{\theta}}{r^{2}}
+\frac{rE^{2}}{2-r}) \nonumber \\
& &  +\frac{1}{2}\frac{\Lambda^{2}L^{2}}{r^{2}\sin^{2}\theta}.
\end{eqnarray}
Similar to Eq. (17), the Hamilton-Jacobi equation is
\begin{eqnarray}
-\Lambda^{2} &=& \frac{r-2}{r} (\frac{\partial S}{\partial r})^{2}+\frac{1}{r^{2}}(\frac{\partial S}{\partial \theta})^{2}
+\frac{rE^{2}}{2-r} \nonumber \\
&& +\frac{\Lambda^{4}L^{2}}{r^{2}\sin^{2}\theta}.
\end{eqnarray}
This equation has no separable form of the variables $r$ and $\theta$ like Eq. (19). That means that
the constant similar to Eq. (20) or (21) is no longer present.
The system (22) holds only one constant (14) and therefore is nonintegrable.
This can be understood easily from the physical point of view. Although the electromagnetic force is absent
for the case of $B\neq0$ with $q=0$, the magnetic field acts as the gravitational effect
and leads to the loss of the second constant in the system. Namely, the gravitational effect of  the magnetic field
included in the Schwarzschild spacetime destroys the integrability of this spacetime.
In this sense, it is possible that chaos is hidden in the system (22).
In the following demonstrations, we focus on the dynamics of order and chaos in this system using numerical
techniques.

\subsubsection{Fourth-order explicit symmetric algorithms in extended phase space}

The conventional fourth-order Runge-Kutta method (RK4) is naturally suitable for explicitly
solving the inseparable Hamiltonian (22).
However, RK4 shows secular growth in the error of the Hamiltonian, as shown in Fig. 3(a).
Instead, symplectic or symmetric methods concern the approximate preservation of
constants of motion without drift in the errors. Usually implicit symplectic
integrators are directly applied for such an inseparable Hamiltonian problem.
A fourth-order implicit symplectic
integrator with a symmetric composition of three second-order symplectic implicit midpoint methods (IS4) [27-29]
exhibits good long term stability and error behavior in Fig. 3(b). In spite of this,
the implicit method needs a lot of additional computational cost compared to the explicit
algorithm RK4. This can be shown clearly in Table 3.

To save the computational cost, Pihajoki [31] doubled the phase space variables and
constructed explicit leapfrog integration schemes of inseparable Hamiltonian systems.
It is good to use two mixing maps on permutations of momenta to restrict the extended (new)
variables to agree with the original (old) ones for given equal initial conditions.
The permutations destroy the symplecticity of the algorithms in general. However,
the methods are symmetric and therefore can still be similar to symplectic integrators
that preserve the original Hamiltonian
without secular growth in the error. In this sense,
these methods are called extended phase space explicit symplectic-like (or symmetric) algorithms.
Liu et al. [32] pointed out that it is better to replace the two permutations of momenta
with the permutations of coordinates and the permutations of momenta (i.e. the sequent two permutations
of coordinates and momenta). Unfortunately, the sequent two permutations
may fail to work in some cases. Instead, the midpoint permutations between  the old variables
and their corresponding new variables are recommended in [33]. In particular,
one of the advantages for the midpoint permutations lies in that the usual symplectic integration
formulae can be directly applied to the extended phase space Hamiltonian but the methods of Pihajoki [31]
and Liu et al. [32] can not. These extended phase space explicit symmetric integrators
have been shown to have good performance in the conservation of the original Hamiltonian
when they are used to solve some inseparable Hamiltonian problems [3,34,35,43].

Now let us consider the application of an extended phase space fourth-order
explicit symmetric algorithm with the midpoint permutations
to the inseparable Hamiltonian (22).
Extending the 4-dimensional phase space variables $(r,\theta, p_{r},p_{\theta})$
to an 8-dimensional phase space variables $(r,\theta, p_{r},p_{\theta};\tilde{r},
\tilde{\theta}, \tilde{p}_{r},\tilde{p}_{\theta})$, we obtain a new
Hamiltonian
\begin{eqnarray}
& & \tilde{H}(r,\theta, p_{r},p_{\theta};\tilde{r}, \tilde{\theta}, \tilde{p}_{r},\tilde{p}_{\theta})
 = H_{1}(r,\theta,\tilde{p}_{r},\tilde{p}_{\theta}) \nonumber \\
& & +H_{2}(\tilde{r}, \tilde{\theta},p_{r},p_{\theta}),
\end{eqnarray}
where $H_{1}=H_{2}=H$, and each of the four pairs $(r,p_{r})$, $(\theta,p_{\theta})$,
$(\tilde{r},\tilde{p}_{r})$ and $(\tilde{\theta},\tilde{p}_{\theta})$ is canonical each other.
The old variables and their corresponding new variables
have the same initial values: $r_{0}=\tilde{r}_{0}$, $\theta_{0}=\tilde{\theta}_{0}$,
$p_{r0}=\tilde{p}_{r0}$ and $p_{\theta0}=\tilde{p}_{\theta0}$.
It is clear that $H_{1}$ and $H_{2}$ are independently, analytically solvable.
Thus $\tilde{H}$ is a separable system with respect to the old variables and the new ones in the extended phase space although
$H$ is not to the old variables.
Taking $\mathbf{H}_{1}$ as an operator for analytically solving the Hamiltonian  $H_{1}$
and  $\mathbf{H}_{2}$ as another operator for analytically solving the Hamiltonian  $H_{2}$,
we have the second-order explicit leapfrog algorithm
\begin{equation}
 S2(h)=\mathbf{H}_{2}(\frac{h}{2})\mathbf{H}_{1}(h)\mathbf{H}_{2}(\frac{h}{2}),
\end{equation}
where $h$ is a time step. A symmetric product of three leapfrogs
yields the fourth-order method of Yoshida [44]
\begin{equation}
 S4(h) = \textbf{M}S2(\lambda_{1}h)S2(\lambda_{2}h)S2(\lambda_{1}h),
\end{equation}
where $\lambda_{1}$ and $\lambda_{2}$ are time coefficients with the following expressions
\begin{equation}
 \lambda_{1}=\frac{1}{2-2^{1/3}}, ~~
 \lambda_{2}=1-2\lambda_{1}.
\end{equation}
Additionally, $\textbf{M}$ is a mixing map on the midpoint permutations between
the old and new variables: $(r+\tilde{r})/2 \rightarrow r$, $(r+\tilde{r})/2 \rightarrow \tilde{r}$,
$(\theta+\tilde{\theta})/2 \rightarrow \theta$, $(\theta+\tilde{\theta})/2 \rightarrow \tilde{\theta}$,
$(p_{r}+\tilde{p}_{r})/2 \rightarrow p_{r}$, $(p_{r}+\tilde{p}_{r})/2 \rightarrow \tilde{p}_{r}$,
$(p_{\theta}+\tilde{p}_{\theta})/2 \rightarrow p_{\theta}$ and $(p_{\theta}+\tilde{p}_{\theta})/2 \rightarrow \tilde{p}_{\theta}$.
As claimed above, this map is powerful to prevent from
the divergence of the numerical solution $(r,\theta,\tilde{p}_{r},\tilde{p}_{\theta})$ for $H_{1}$
and the numerical solution $(\tilde{r},\tilde{\theta},p_{r},p_{\theta})$ for $H_{2}$ [or the
original solution $(r, \theta, p_{r}, p_{\theta})$
and the extended solution $(\tilde{r},\tilde{\theta},\tilde{p}_{r},\tilde{p}_{\theta})$] with time.
As a result, the extended phase space explicit symmetric algorithm S4 in Fig. 3(c) gives excellent behavior in
the error growth and conservation of the original Hamiltonian.
In view of such good computational efficiency and accuracy, we employ the method S4 to
investigate the dynamics of neutral particles.

\subsubsection{Chaotic indicators}

Since the system (22) has four dimensions and the integral (14),
its structure in the original phase space
can be described clearly with the aid of Poincar\'{e} section method.
When the energy of the system (22) is not too high, e.g. $E=0.99$ in Fig. 4(a),
all phase orbits on the section at the plane $\theta=\pi/2$ are KAM tori
and therefore are quasi-periodic, regular.
When the magnetic field $B=0.001$ is fixed but the energy slightly increases,
e.g. $E=0.9905$, one of the orbits
is no longer a torus and has a small number of random distributed points in Fig. 4(b).
This seems to show the chaoticity of this orbit.
With a further increase of  the energy, e.g. $E=0.9915$, two orbits
have a large number of random distributed points
in Fig. 4(c). That seems to mean that the extent of chaos is typically strengthened.

In fact, the orbital dynamical feature of order and chaos depends on
the rate of divergence of an orbit and its nearby orbit in the phase space.
The rate can be measured precisely  by the  maximum Lyapunov exponent [36,37]
\begin{eqnarray}
\lambda=\lim_{\tau\rightarrow\infty}\frac{1}{\tau}\ln\frac{d(\tau)}{d(0)}.
\end{eqnarray}
Here, $d(\tau)$ is the proper distance between the two nearby orbits at the proper time $\tau$
and is defined by
\begin{eqnarray}
d(\tau)=\sqrt{|g_{\alpha\beta}\Delta x^{\alpha}\Delta x^{\beta}|}.
\end{eqnarray}
$\Delta x^{\alpha}=(\Delta t, \Delta r, \Delta \theta, \Delta \phi)$
denotes a separation vector between the two nearby orbits, and $(\Delta t, \Delta \phi)$
can be obtained by integrating Eq. (11) with Eq. (13).
$d(0)$ is the initial distance. This Lyapunov exponent  is independent of the choice of
time and space coordinates. In this sense, it is called as an invariant chaotic indicator.
Fig. 4(d) plots
the Lyapunov exponents of all orbits in Fig. 4(a).
Because these exponents do not remain stabilizing values and seem to decrease to zero with time  when
the time $\tau=10^{6}$, all orbits are typically regular.
The Lyapunov exponent of one of the bounded orbits in Fig. 4(b) tends to a stabilizing positive value
in Fig. 4(e) and shows the chaoticity of this orbit.
It is easy to find that the two bounded orbits in Fig. 4(c) have positive Lyapunov exponents in Fig. 4(f), and should be chaotic.
Clearly, the non-zero positive Lyapunov exponents of the two orbits in Fig. 4(f) are larger than
that of the orbit in Fig. 4(e). Thus, chaos in  Fig. 4(f) becomes stronger than in  Fig. 4(e).
These results obtained from the method of Lyapunov exponents are completely consistent with those shown via
the method of Poincar\'{e} sections.

It is in sufficient long times that the Lyapunov exponents of these chaotic orbits reach their stabilizing values.
Are there any methods that are quicker to find chaos than the Lyapunov exponents? Yes, there are.
A fast Lyapunov indicator (FLI) is an example of them.
It is calculated according to the following form [37]
\begin{equation}
FLI=\log_{10}\frac{d(\tau)}{d(0)}.
\end{equation}
This indicator remains invariant when different
time and space coordinates are used.
If the FLI of a bounded orbit increases exponentially with time $\log_{10} \tau$, then
this orbit is chaotic; if it grows in a power law, this orbit should be ordered.
Such dramatic different variations of the FLIs are used to distinguish between the chaotic and regular
two cases. Based on this criterion, the FLIs in an integration time
of $\tau=5\times 10^{4}$ in Fig. 4(g) show that none of the orbits in Fig. 4(a) is chaotic.
Seen from the FLIs  in Fig. 4(h), only one of  the orbits in Fig. 4(b) behaves chaotic behavior.
In particular, the FLIs of the two chaotic orbits in Fig. 4(i) grow faster than
the FLI of the chaotic orbit in Fig. 4(h). This means that chaos in Fig. 4(c) is stronger than in
Fig. 4(b).

The methods of Poincar\'{e} sections, Lyapunov exponents and FLIs
provided the consistent result that chaos can occur in the  magnetized Ernst-Schwarzschild spacetime
although these particles without charges have no interaction of the electromagnetic force.
This chaoticity is caused due to
the magnetic field acting as the gravitational effect and destroying the
integrability of the original spacetime.
These methods also showed that chaos is strengthened typically with
the energy increasing. In a similar way, we
can confirm that the chaotic behavior occurs more easily  when the magnetic field $B$ increases for a given energy
(e.g. $E=0.991$) in Fig. 5.
This is because an increase of the magnetic field (or the energy) means that of the gravitational effect from the magnetic field.
The integrability of the original spacetime is typically destroyed so that there is no great difference between
the gravitational effect of the magnetic field and the gravity of the black hole.

\subsection{Case 3: $B\neq0$, $q\neq0$}

For the case of $B\neq0$ with $q\neq0$, charged particles move in the electromagnetic field and spacetime geometry.
Without doubt, the system (12) for the description of the motion of charged particles
is nonintegrable. Thus, the onset of chaos is not unexpected.

The three types of parameter combinations in Fig. 5 are still considered, but the charge $q=0.01$ is chosen in Fig. 6.
Although the electromagnetic force is included, the phase space structures for each parameter combination in Fig. 6 are not
explicitly different from those for the same  parameter combination in Fig. 5.
An increase of the magnetic field still leads to that of chaos. When the charge $q=-0.01$ is used instead of the charge $q=0.01$,
no typical changes of the phase space structures occur in Fig. 7.

Fixing the parameters $E=0.9913$, $L=3.8$ and $B=0.001$, we find in Fig. 8 that a larger charge results in stronger chaos.
If the fixed magnetic field $B=-0.001$ is used in Fig. 9, increasing the charge gives rise to
decreasing the extent of chaos. These facts show that the electromagnetic forces in Figs. 8 and 9 exert
different influences on chaos. Even if the electromagnetic forces are much small in Figs. 8(a) and 9(a),
weak chaos is present. As mentioned above, this chaotic behavior is mainly due to
the gravitational effect of the magnetic field.
With the electromagnetic forces increasing,  the electromagnetic forces in Fig. 8 (b) and (c) strengthen
the extent of chaos caused by the gravitational effect of the magnetic field, whereas
those in Fig. 9 (b) and (c) suppress or weaken the extent of chaos.

\section{Summary}

In this work, we provide some insight into the dynamics of both electrically charged and
neutral particles around the Schwarzschild black hole in the magnetic universe.
The effective potentials and stable circular orbits at the equatorial plane are discussed.
For weak magnetic or electromagnetic fields,  the radii of the innermost stable circular orbits approach to that of the Schwarzschild spacetime.
If these particles have no charges and the electromagnetic forces are absent, then the magnetic field acts as the gravitational effect and destroys the
integrability of the original spacetime. Because of this, the occurrence of chaos is possible.
In some cases, chaos is strengthened typically with
the energy or the magnetic field increasing. When the charged particles move in the electromagnetic field and spacetime geometry,
the electromagnetic forces play an important role in strengthening or suppressing
the extent of chaos caused by the gravitational effect of the magnetic field.

\section*{Acknowledgments}

This research has been supported by the National
Natural Science Foundation of China under Grant No. 11533004 and the Natural Science
Foundation of Jiangxi Province under Grant No. 20153BCB22001.


\newpage

\begin{table*}
\begin{center}
\small \caption{Radii $r$ of stable circular orbits at the equatorial plane in Fig. 1.}
\label{Table 1}
\begin{tabular}{cccccccc}
\hline
       q        & $B$           & {L}       & {r}               & q             & $B$             & {L}       & {r}            \\
\hline
  \quad         & \quad         & 3.4641    & 6                 & \quad         & \quad           & 3.4643    & 5.999136705541  \\
       0        & 0             & 3.8643    & 10.77999999999    &    0          & $10^{-3}$    & 3.8643    & 10.76999981515  \\
  \quad         & \quad         & 4.1       & 12.89999999999    & \quad         & \quad           & 4.1       & 12.87999976799  \\

\hline
  \quad         & \quad          & 3.4643    & 5.999136706604   & \quad         & \quad           & 3.4643    & 5.999136703616 \\
  $\pm10^{-3}$  & $\pm10^{-3}$   & 3.8543    & 10.66999981739   & $\pm10^{-3}$  & $\mp10^{-3}$    & 3.8356    & 10.49999982119 \\
   \quad        &  \quad         & 4.1       & 12.87999976799   & \quad         & \quad           & 4.1       & 12.87999976799 \\
\hline
  \quad          & \quad           & 3.4637  & 5.999132540466    & \quad        & \quad           & 3.4649    & 5.999132244625 \\
  $\pm0.1$       & $\pm10^{-3}$    & 3.8345  & 10.48999982141    & $\pm0.1$     & $\mp10^{-3}$    & 3.8345    & 10.49999982119  \\
  \quad          & \quad           & 4.1     & 12.86999976821    & \quad        & \quad           & 4.1       & 12.88999976777  \\
\hline
\end{tabular}
\end{center}
\end{table*}

\begin{table*}
\begin{center}
\small \caption{Radii $r$ of the innermost stable circular orbits at the equatorial plane in Fig. 2.}
\label{Table 2}
\begin{tabular}{cccccc}
\hline
       q         & ${B}$                      & {r}              & q              & ${B}$                      & {r}             \\
\hline
       0         & 0                          & 6                & 0              & $10^{-3}$               & 5.999136705541   \\
\hline
  $\pm10^{-4}$   & $\pm10^{-3}$            & 5.999136705686      & $\pm10^{-4}$   & $\mp10^{-3}$               & 5.999136705382   \\
\hline
  $\pm10^{-3}$   & $\pm10^{-3}$            & 5.999136706604      & $\pm10^{-3}$   & $\mp10^{-3}$               & 5.999136703616   \\
\hline
  $\pm10^{-2}$   & $\pm10^{-3}$            & 5.999136677351      & $\pm10^{-2}$   & $\mp10^{-3}$               & 5.999136647471   \\
\hline
  $\pm0.1$       & $\pm10^{-3}$            & 5.999132244625      & $\pm0.1$       & $\mp10^{-3}$               & 5.999132540466   \\
\hline
  $\pm1$         & $\pm10^{-3}$            & 5.998705579809      & $\pm1$         & $\mp10^{-3}$               & 5.998705579809   \\

\hline
  $\pm10^{-3}$   & $\pm10^{-2}$            & 5.919971855034      & $\pm10^{-3}$   & $\mp10^{-2}$               & 5.919969263988   \\
\hline
  $\pm10^{-2}$   & $\pm10^{-2}$            & 5.919979831736      & $\pm10^{-2}$   & $\mp10^{-2}$               & 5.919953924028   \\
\hline
  $\pm0.1$       & $\pm10^{-2}$            & 5.919727012147      & $\pm0.1$       & $\mp10^{-2}$               & 5.919470675229   \\
\hline
  $\pm1$         & $\pm10^{-2}$            & 5.884025383442      & $\pm1$         & $\mp10^{-2}$               & 5.884025383442   \\
\hline
\end{tabular}
\end{center}
\end{table*}

\begin{table*}
\begin{center}
\small \caption{CPU times (seconds) for the three algorithms in Fig. 3.}
\label{Table 3}
\begin{tabular}{cccc}\hline
 Algorithm   & RK4 & IS4  & S4 \\
\hline
CPU  Time     & 10  & 50   & 27 \\
\hline
\end{tabular}
\end{center}
\end{table*}

\newpage

\begin{figure*}
\center{
\includegraphics[scale=0.2]{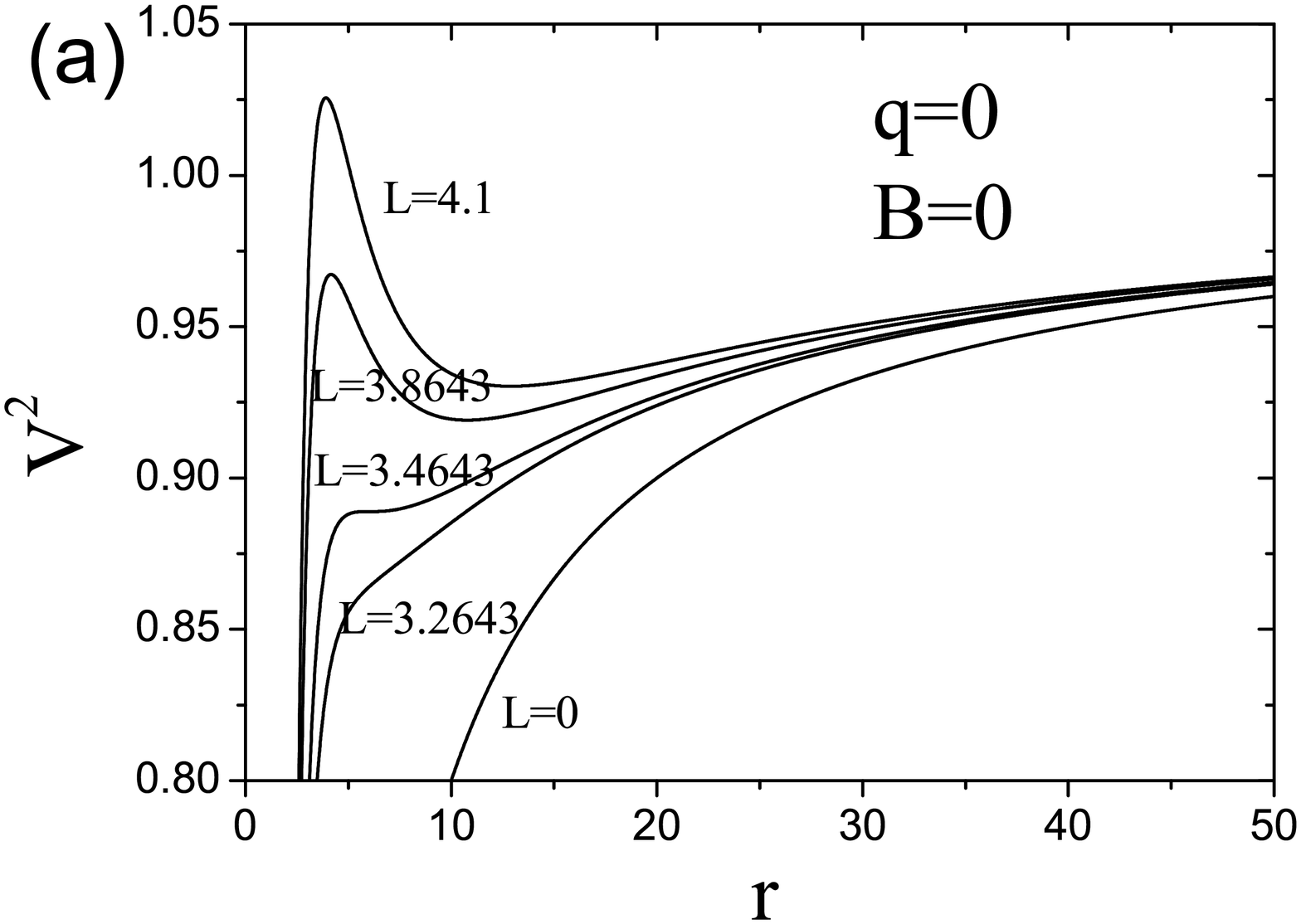}
\includegraphics[scale=0.2]{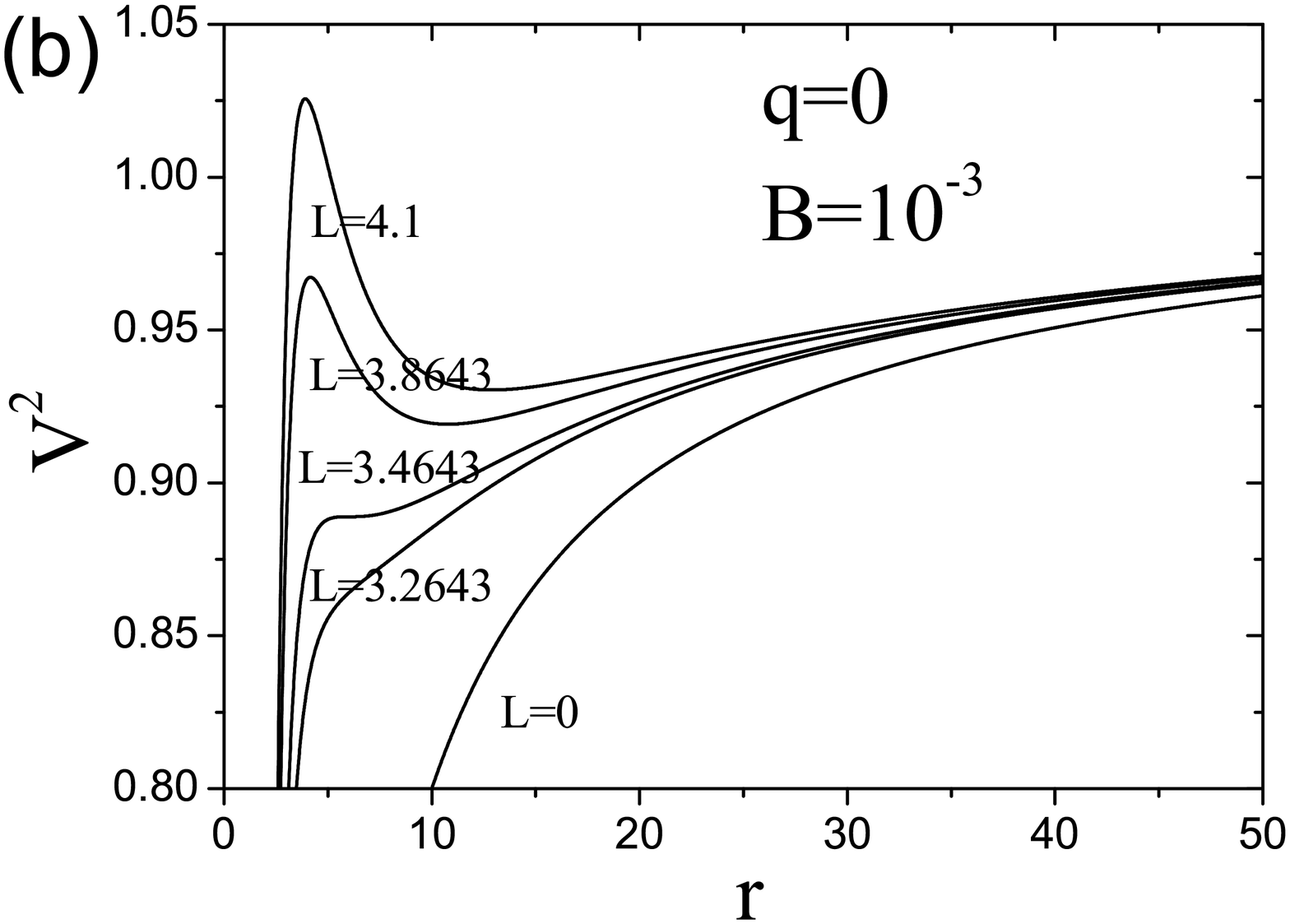}
\includegraphics[scale=0.2]{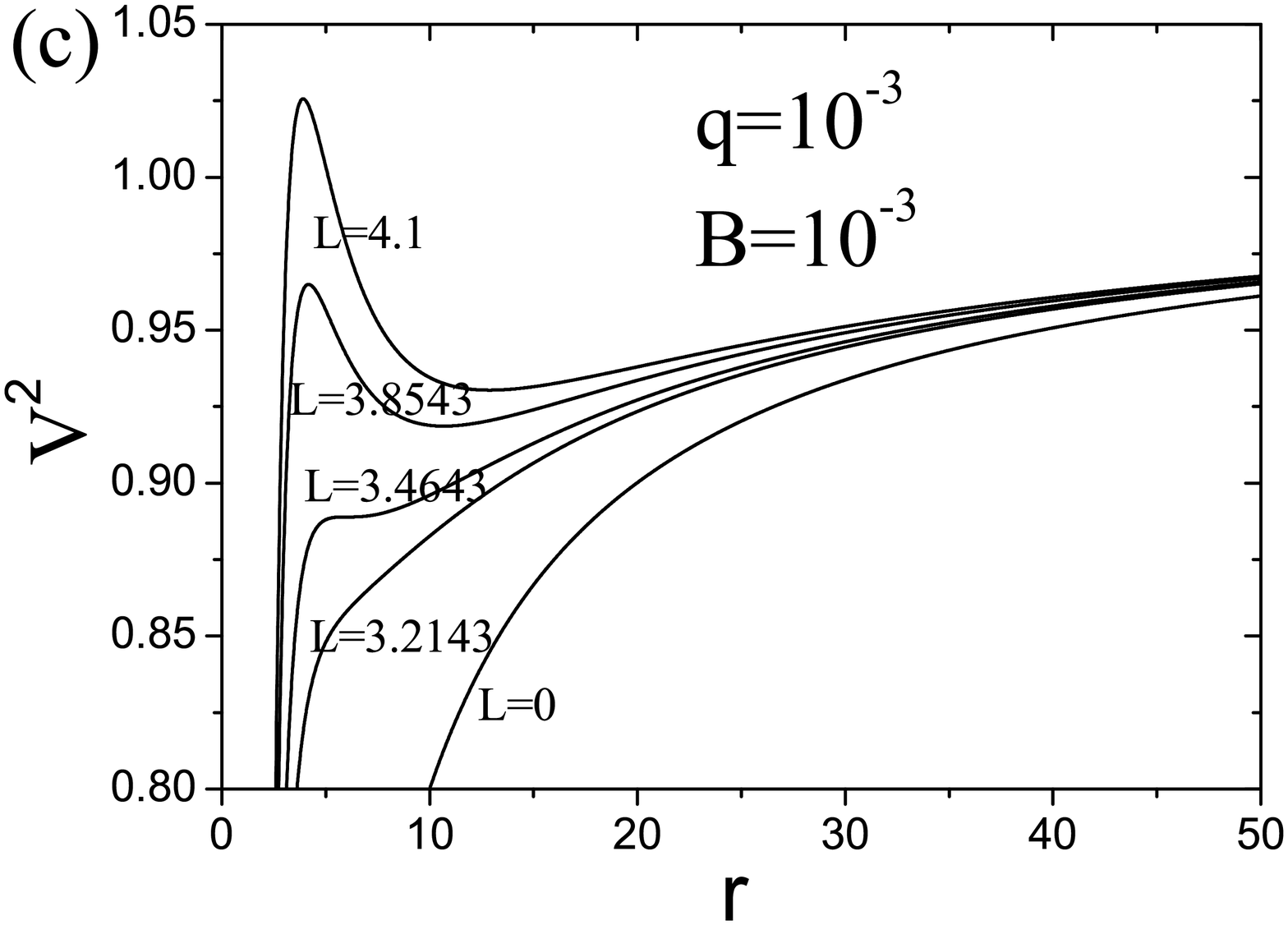}
\includegraphics[scale=0.2]{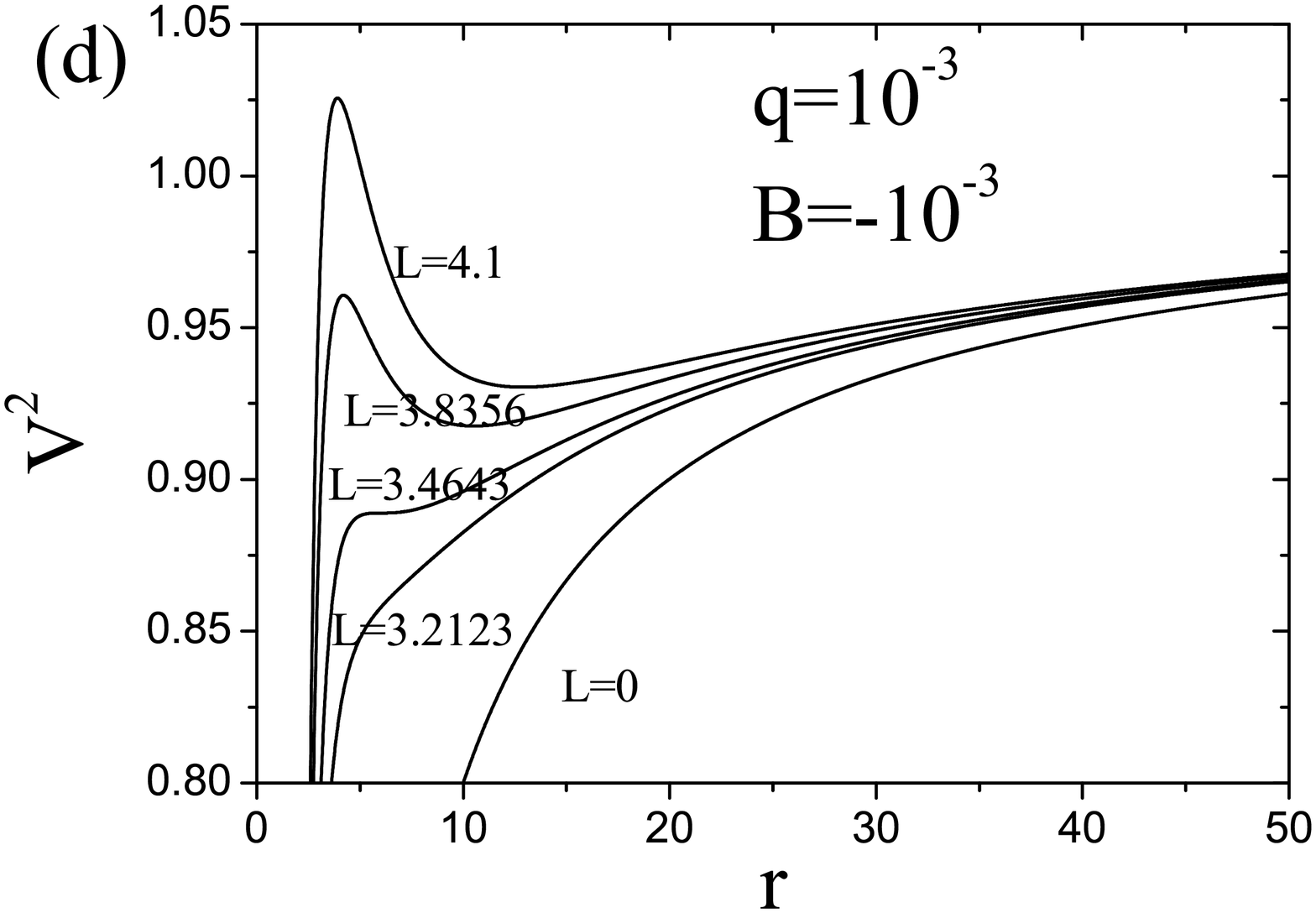}
\includegraphics[scale=0.2]{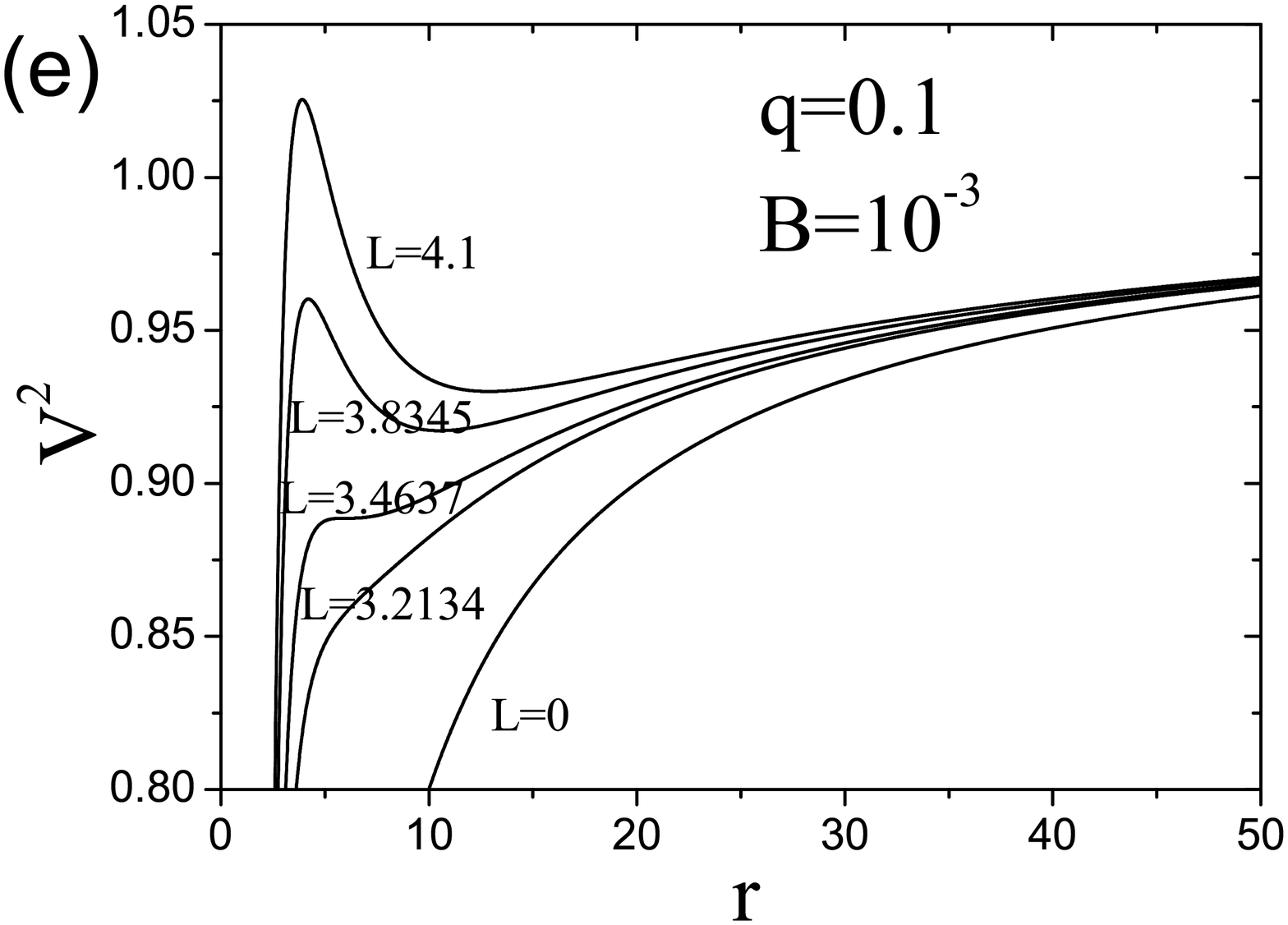}
\includegraphics[scale=0.2]{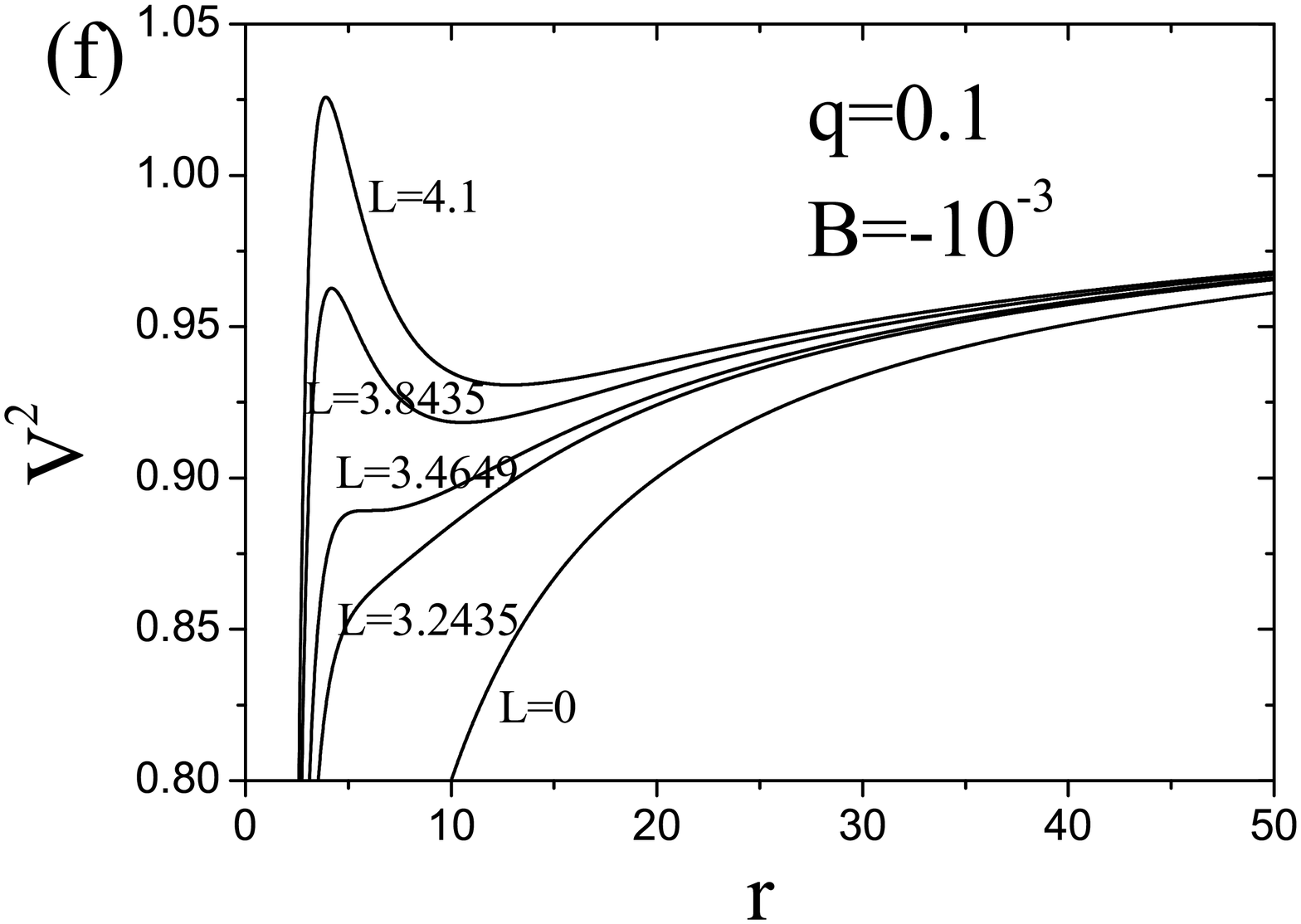}
\caption{Effective potentials for several sets of parameter combinations.
}} \label{fig1}
\end{figure*}

\begin{figure*}
\center{
\includegraphics[scale=0.5]{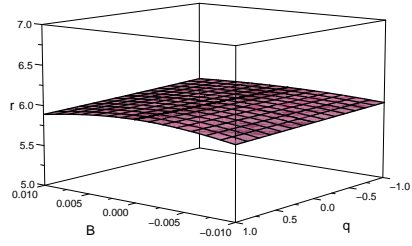}
\caption{(color online) Radii $r$ of the innermost stable circular orbits varying with
 the parameters $B$ and $q$.
}} \label{fig2}
\end{figure*}

\begin{figure*}
\center{
\includegraphics[scale=0.2]{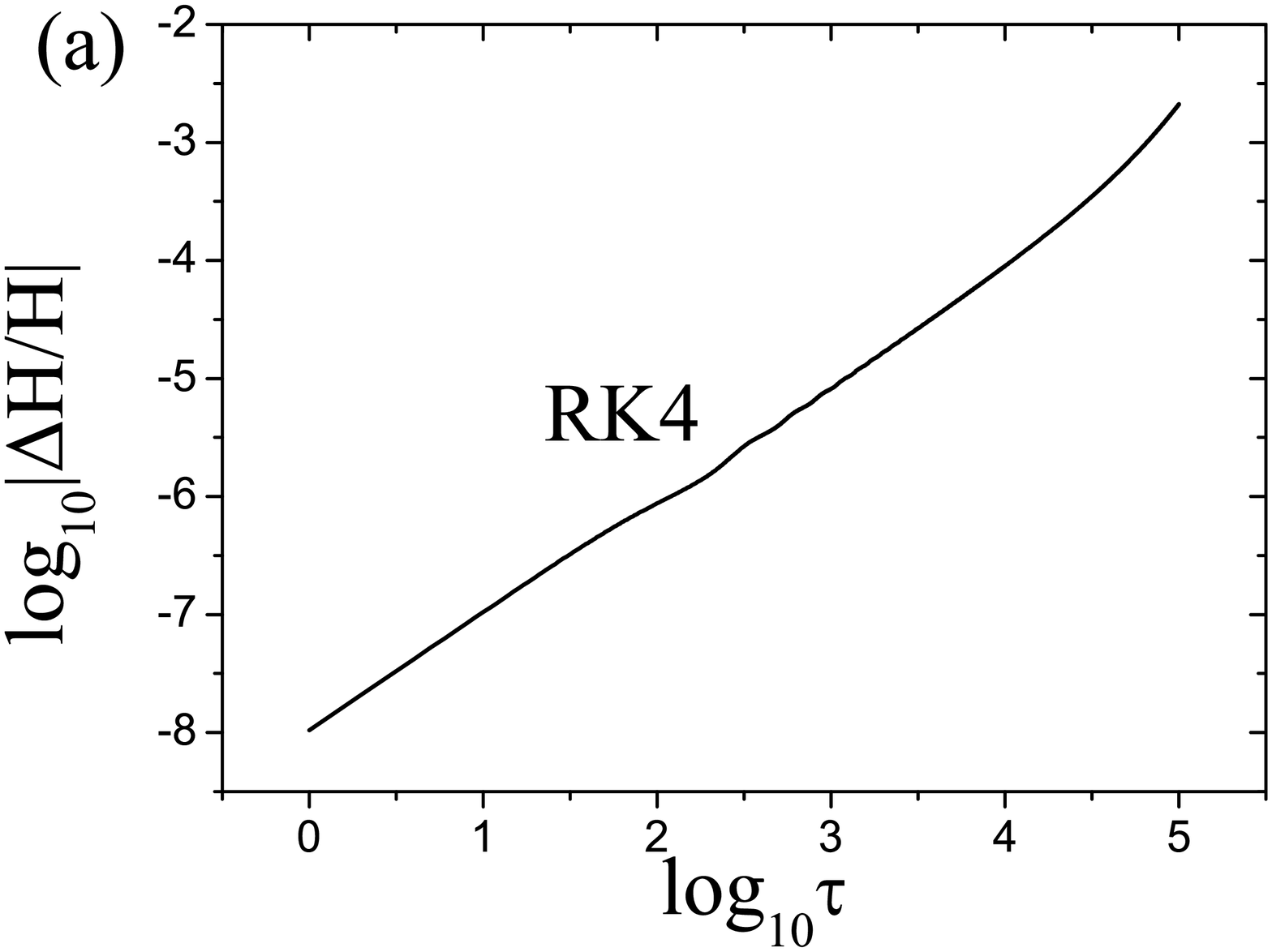}
\includegraphics[scale=0.2]{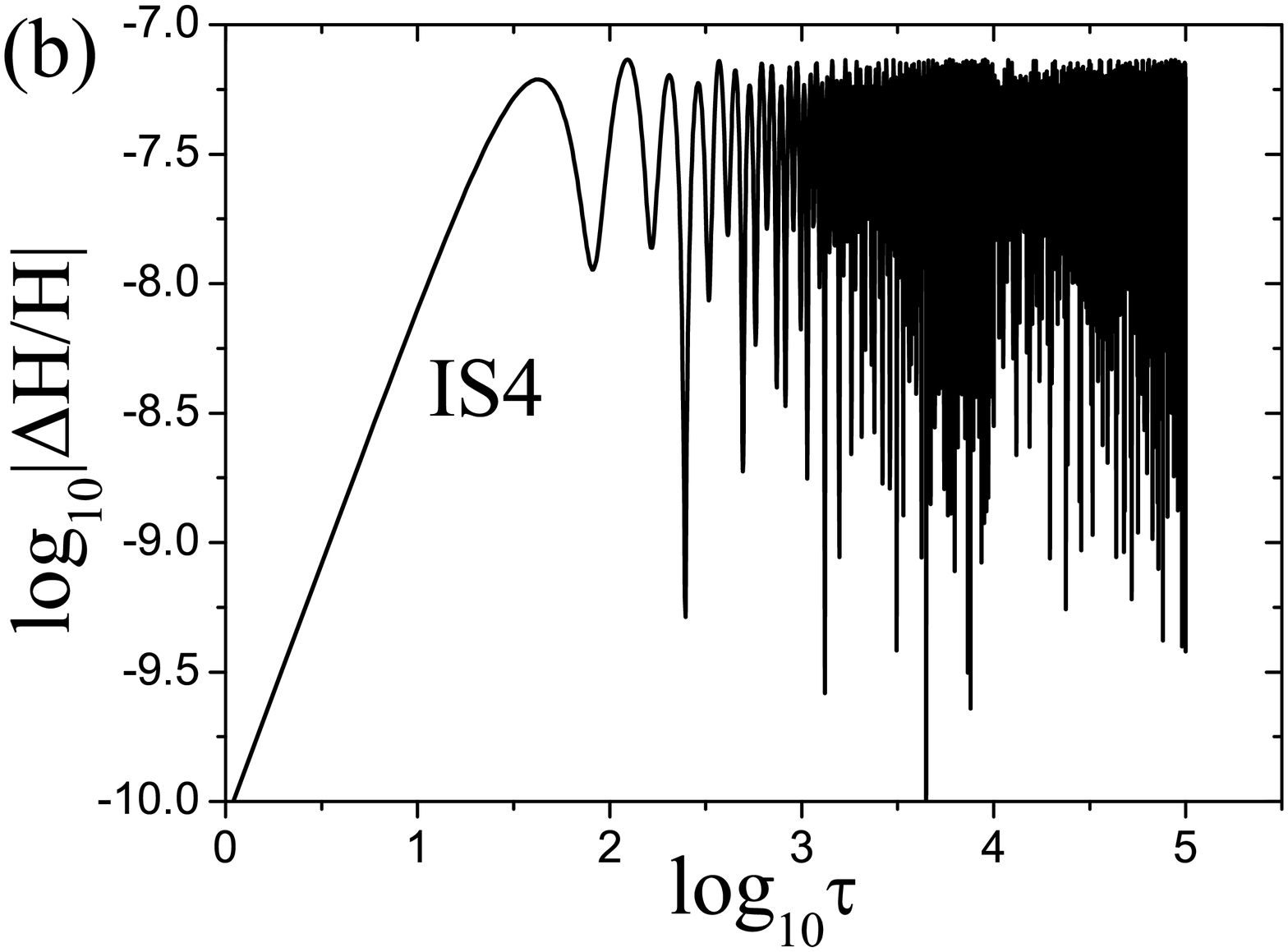}
\includegraphics[scale=0.2]{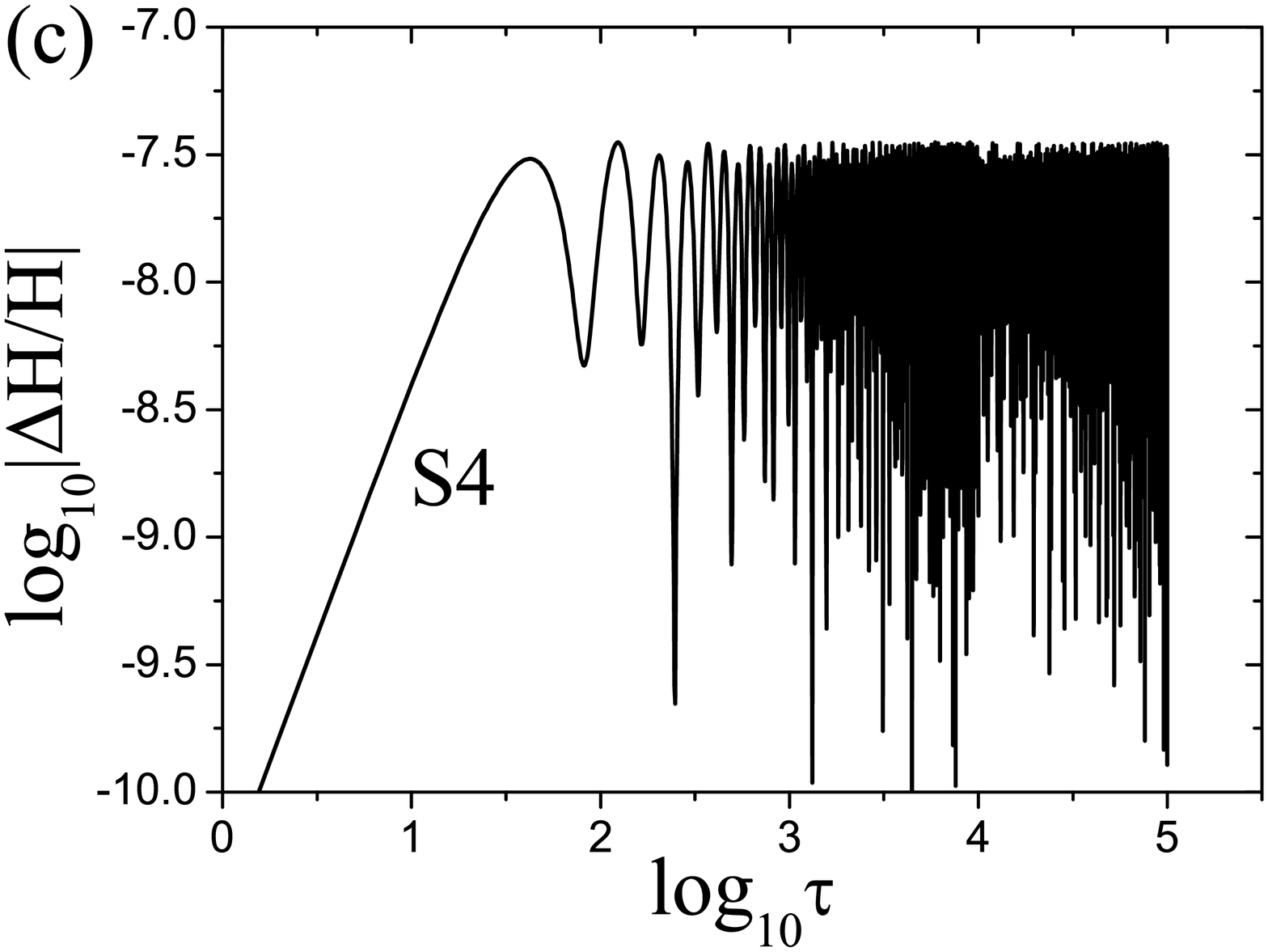}
\caption{Relative errors of the original Hamiltonian (22) for three integrators, including
 the conventional fourth-order explicit Runge-Kutta method RK4, the fourth-order implicit symplectic integrator IS4
 and the extended phase space fourth-order explicit symmetric method S4. The time step is $h=0.1$, and the parameters
 are $q=0$,  $B=10^{-4}$, $E=0.956$ and $L=3.6$. The initial conditions
are $r=10$, $p_{r}=0$, $\theta=\pi/2$, and the initial value of $p_{\theta}>0$ is determined by Eq. (14).
}} \label{fig3}
\end{figure*}

\begin{figure*}
\center{
\includegraphics[scale=0.2]{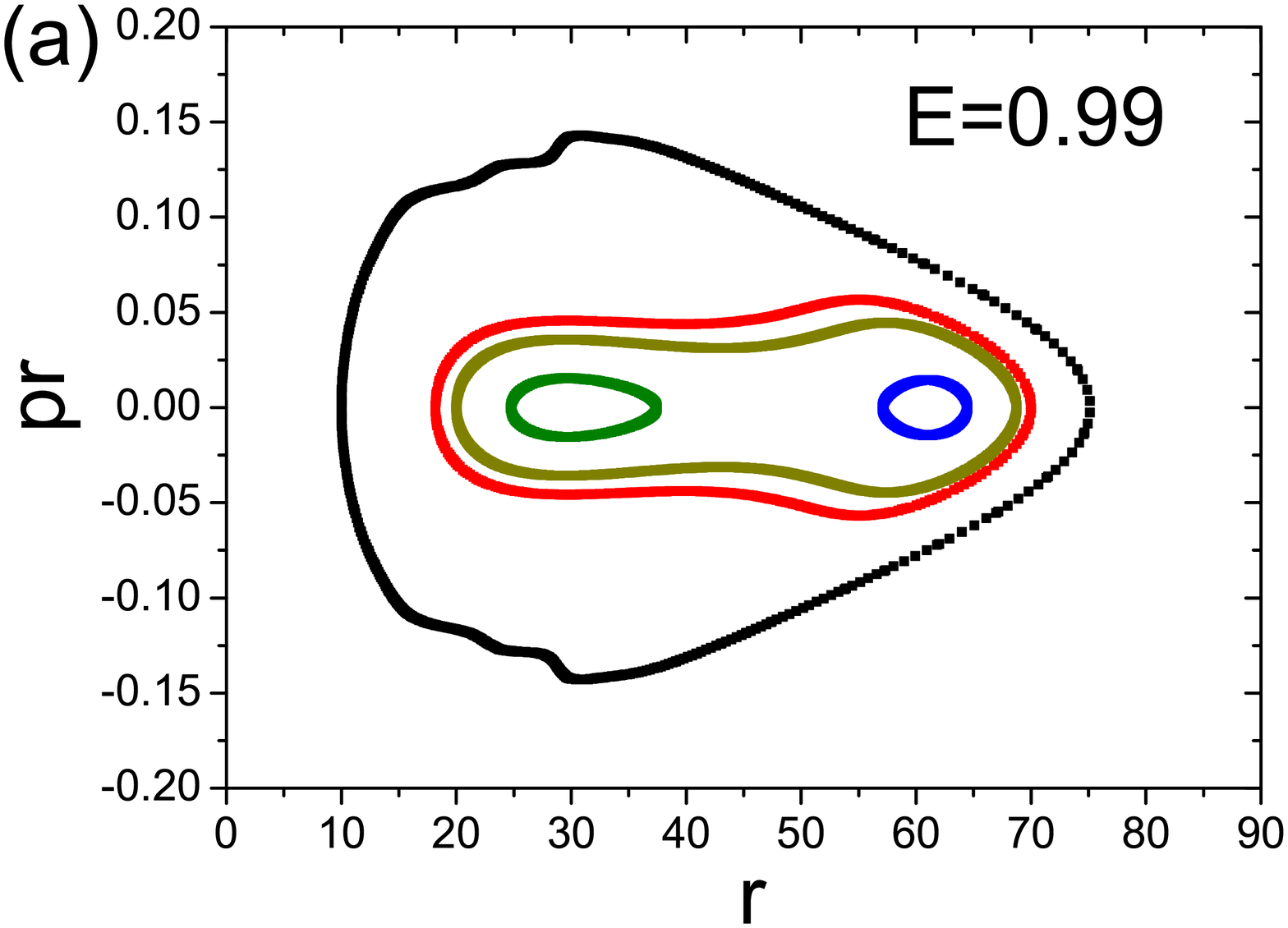}
\includegraphics[scale=0.2]{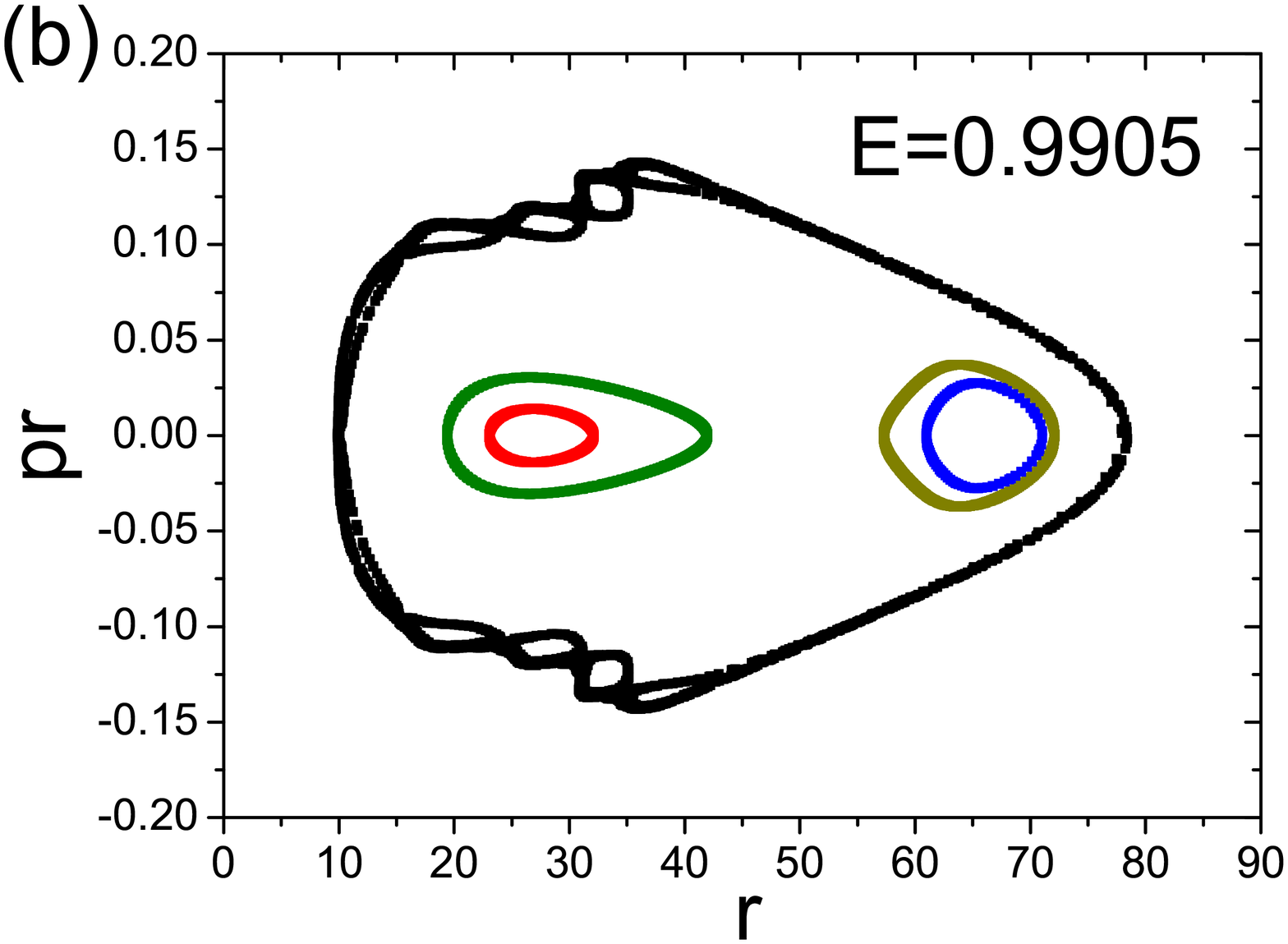}
\includegraphics[scale=0.2]{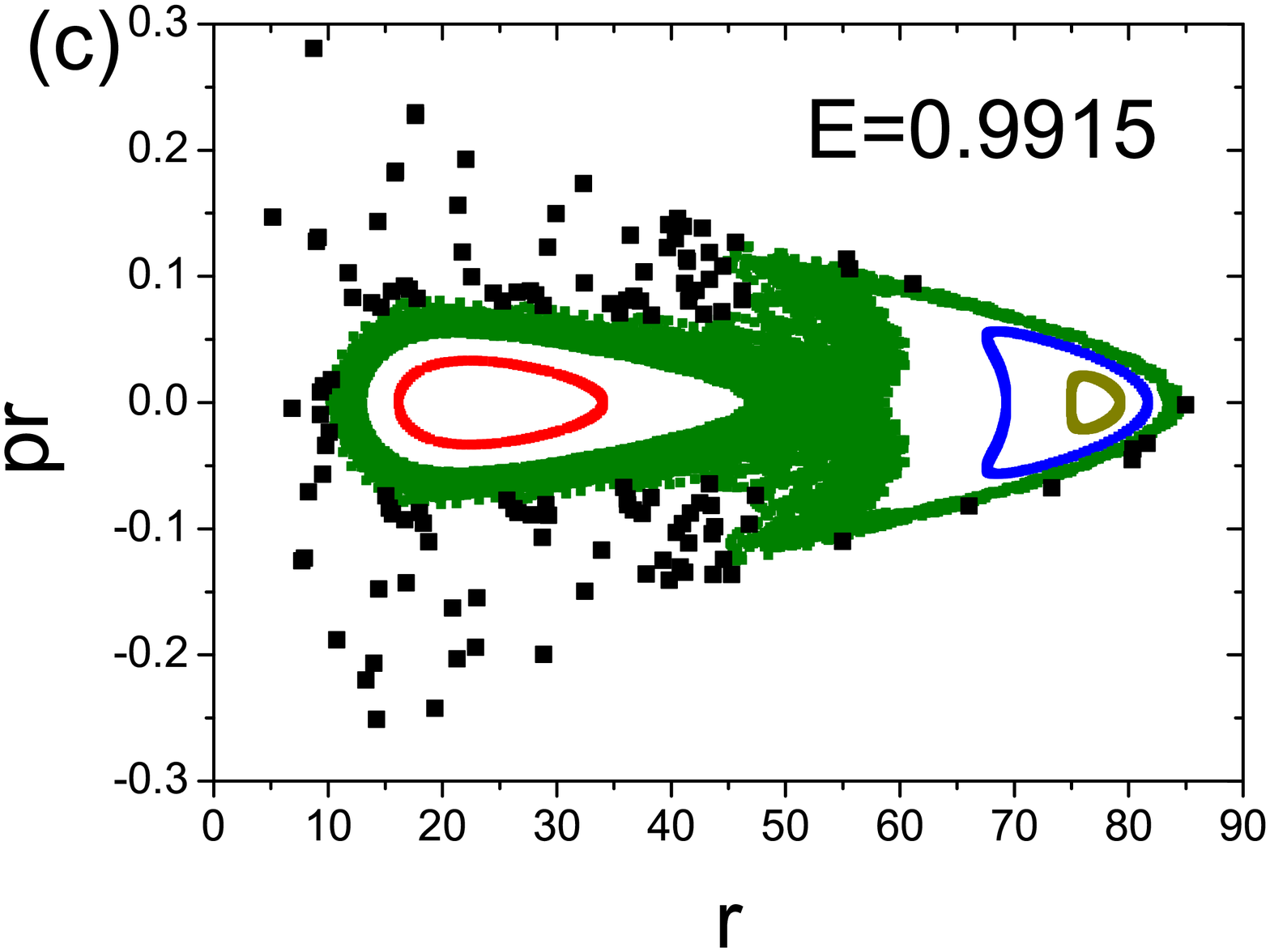}
\includegraphics[scale=0.2]{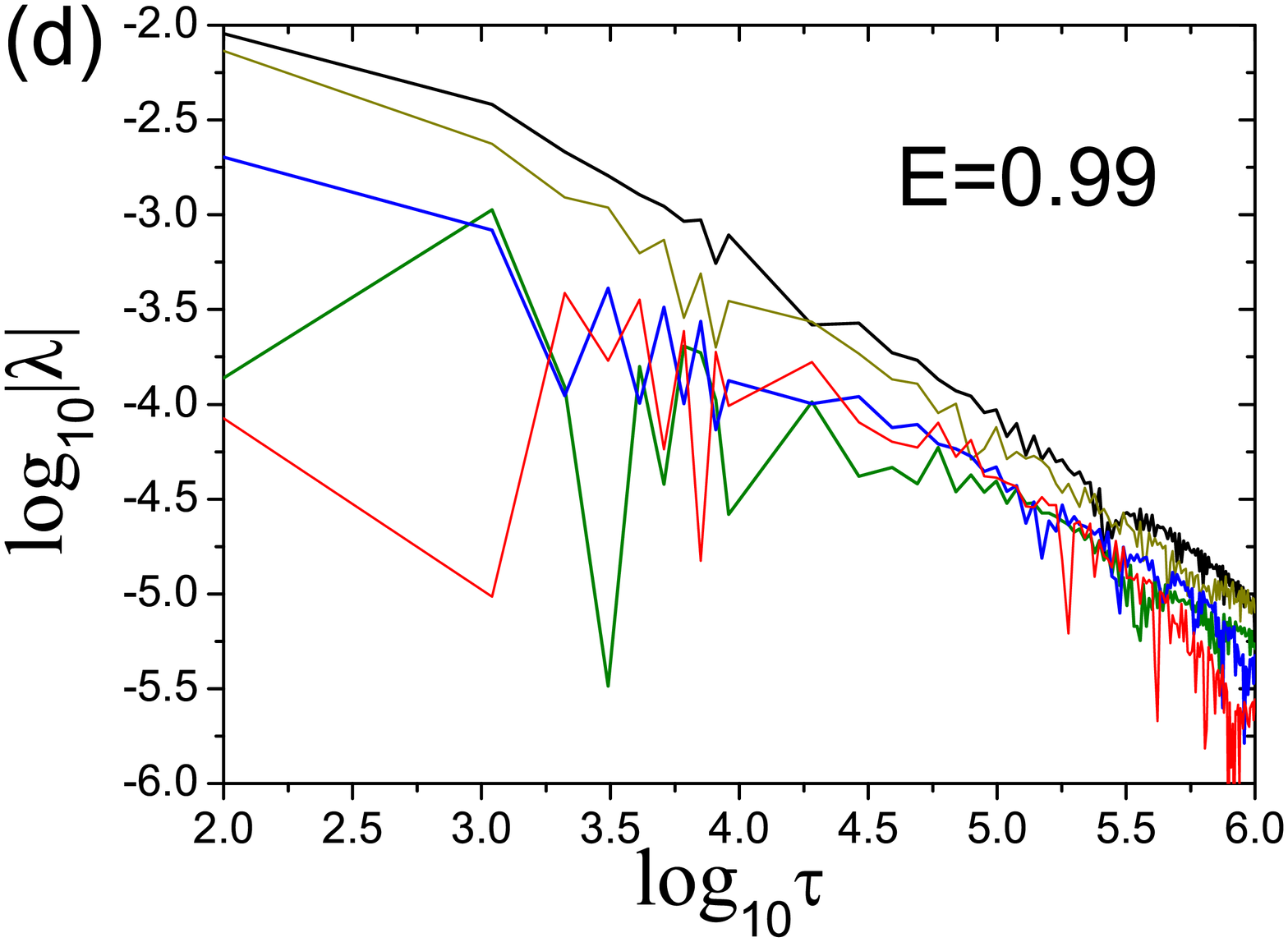}
\includegraphics[scale=0.2]{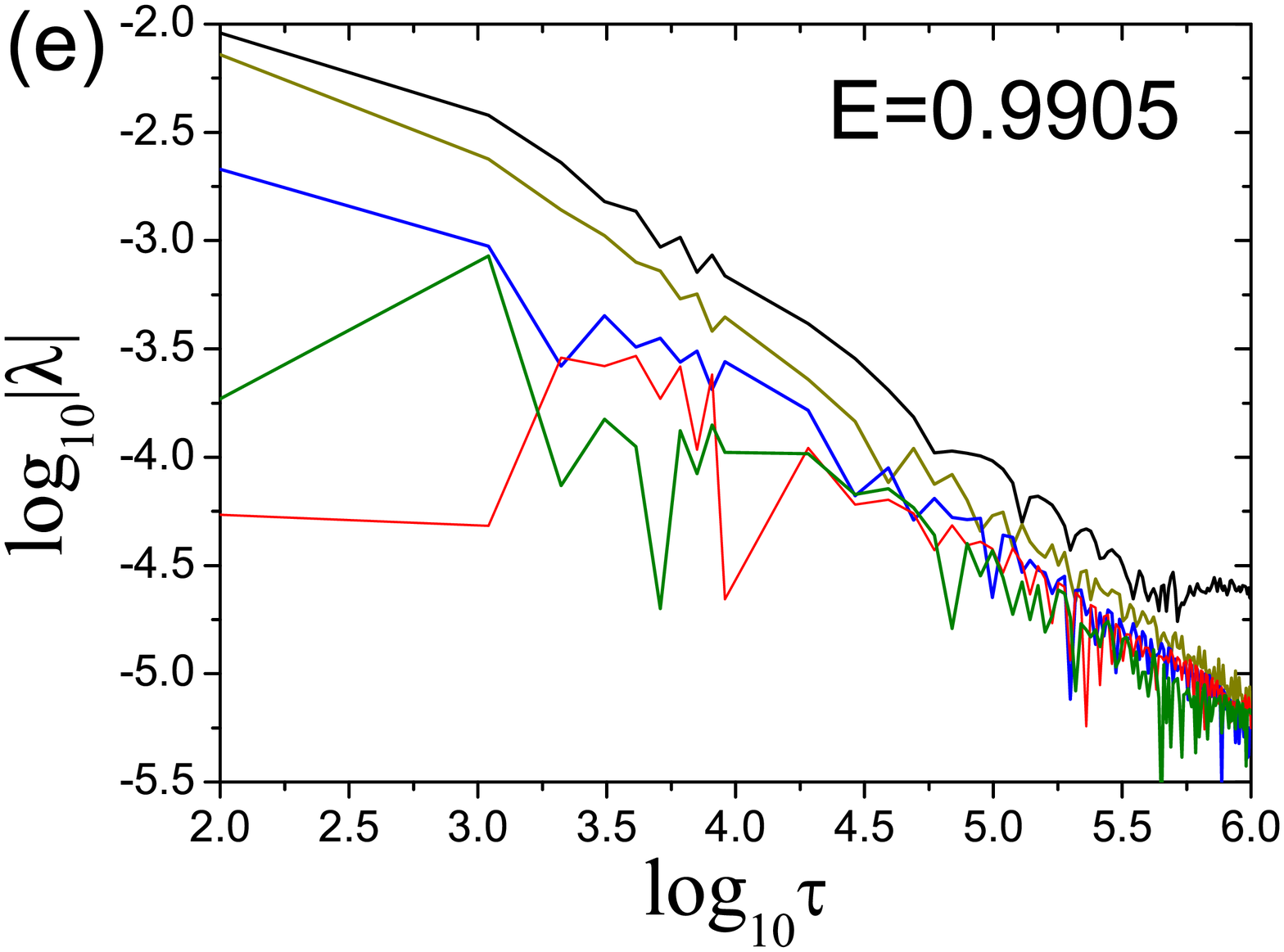}
\includegraphics[scale=0.2]{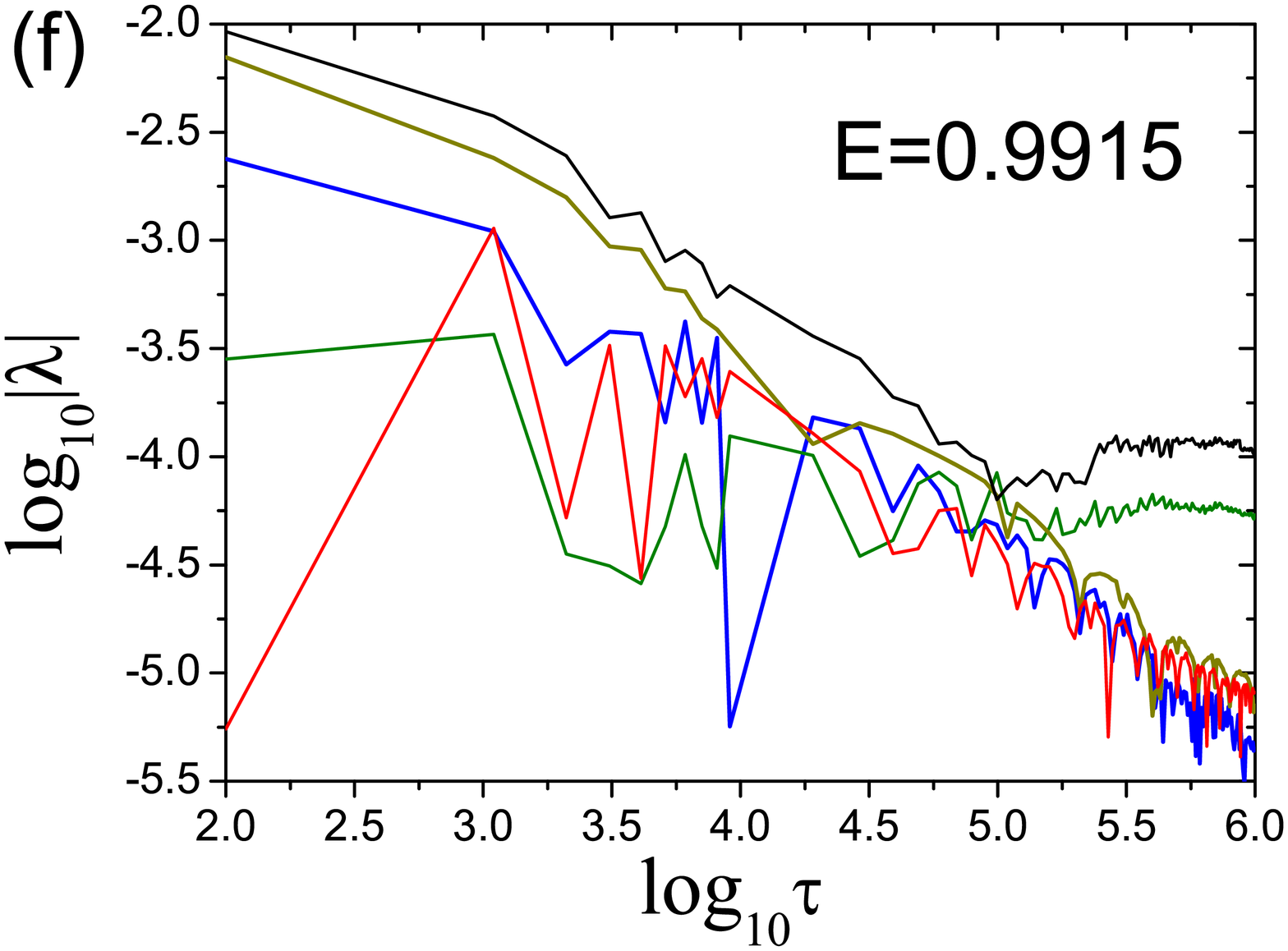}
\includegraphics[scale=0.2]{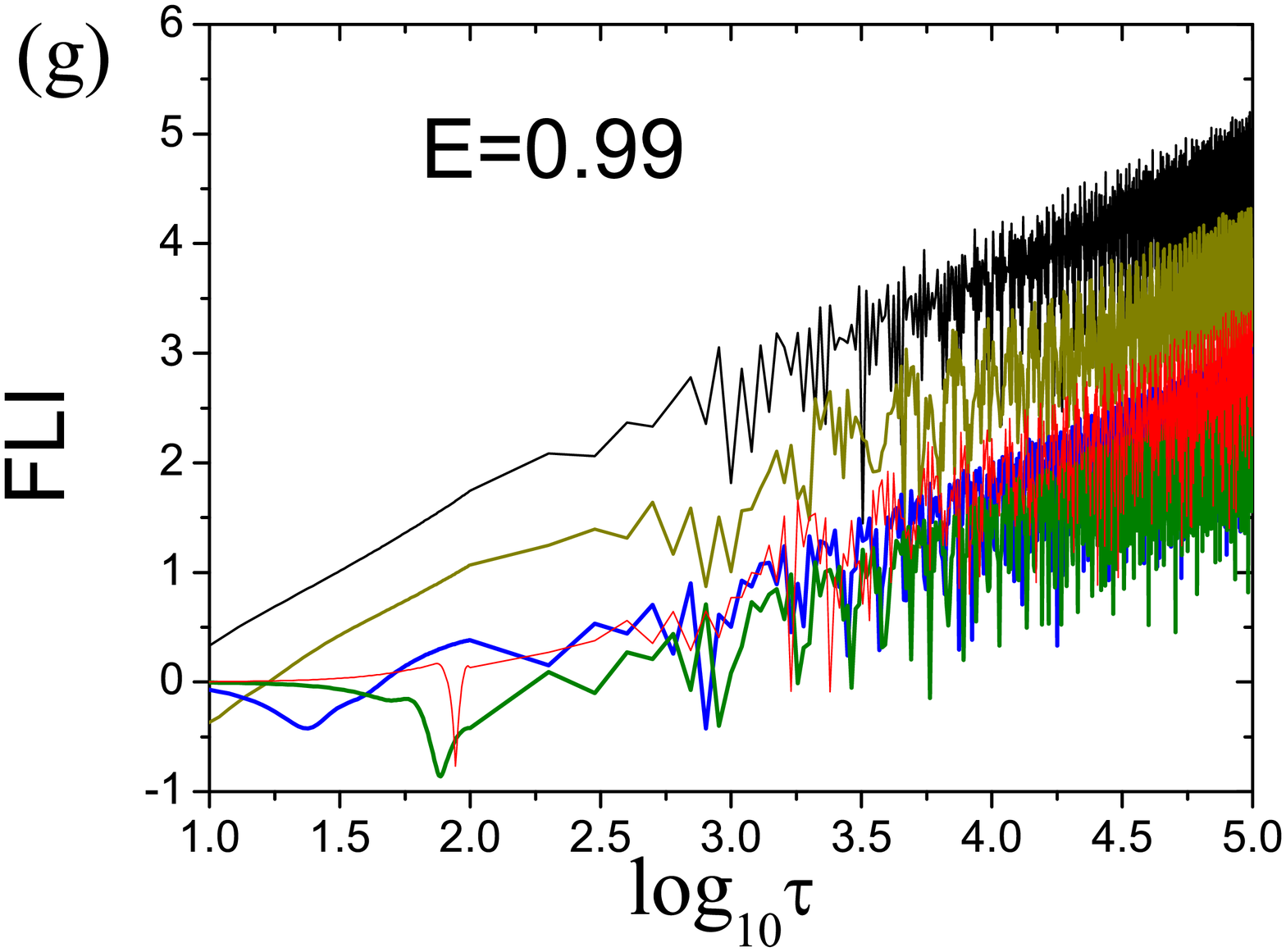}
\includegraphics[scale=0.2]{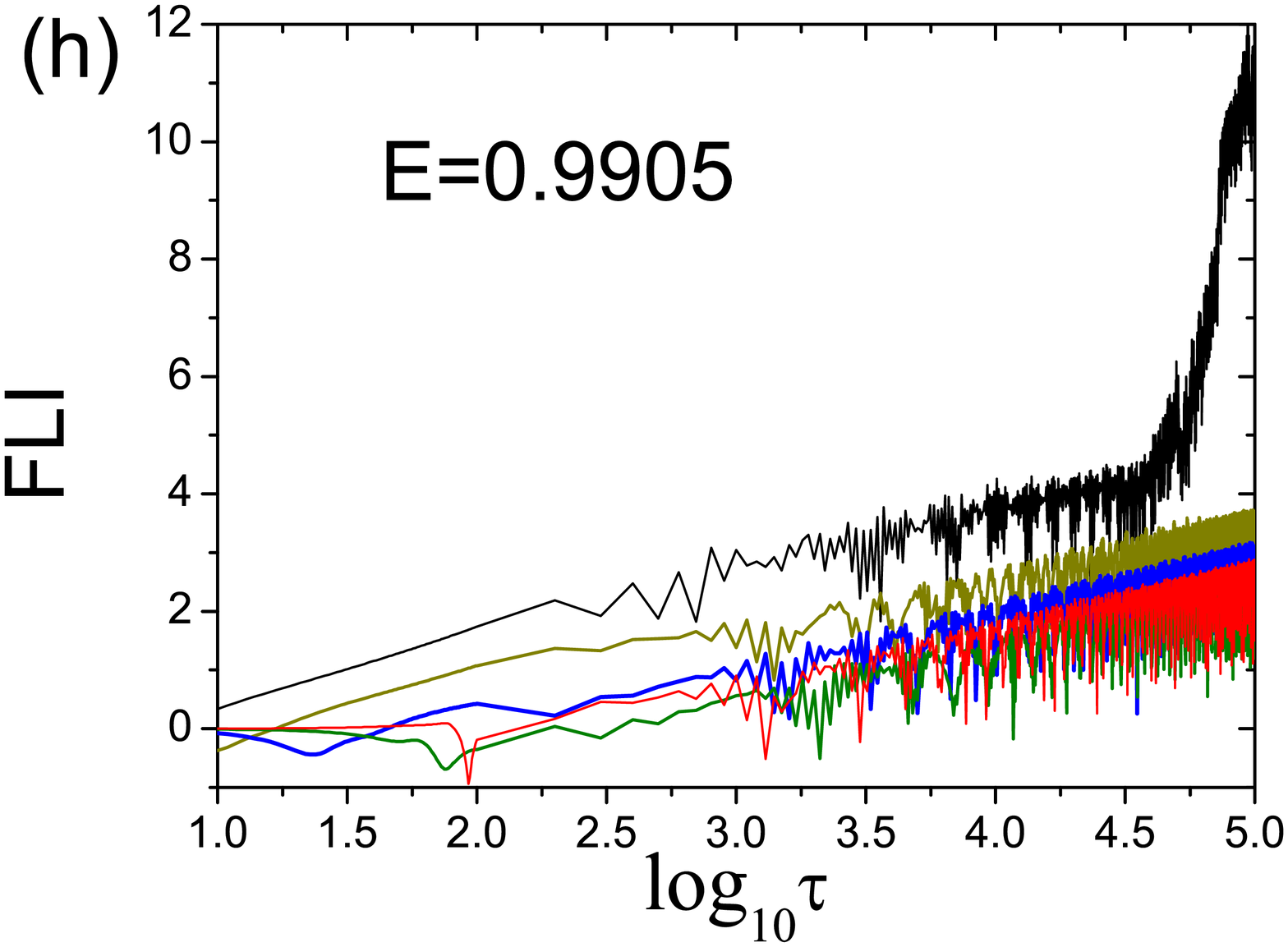}
\includegraphics[scale=0.2]{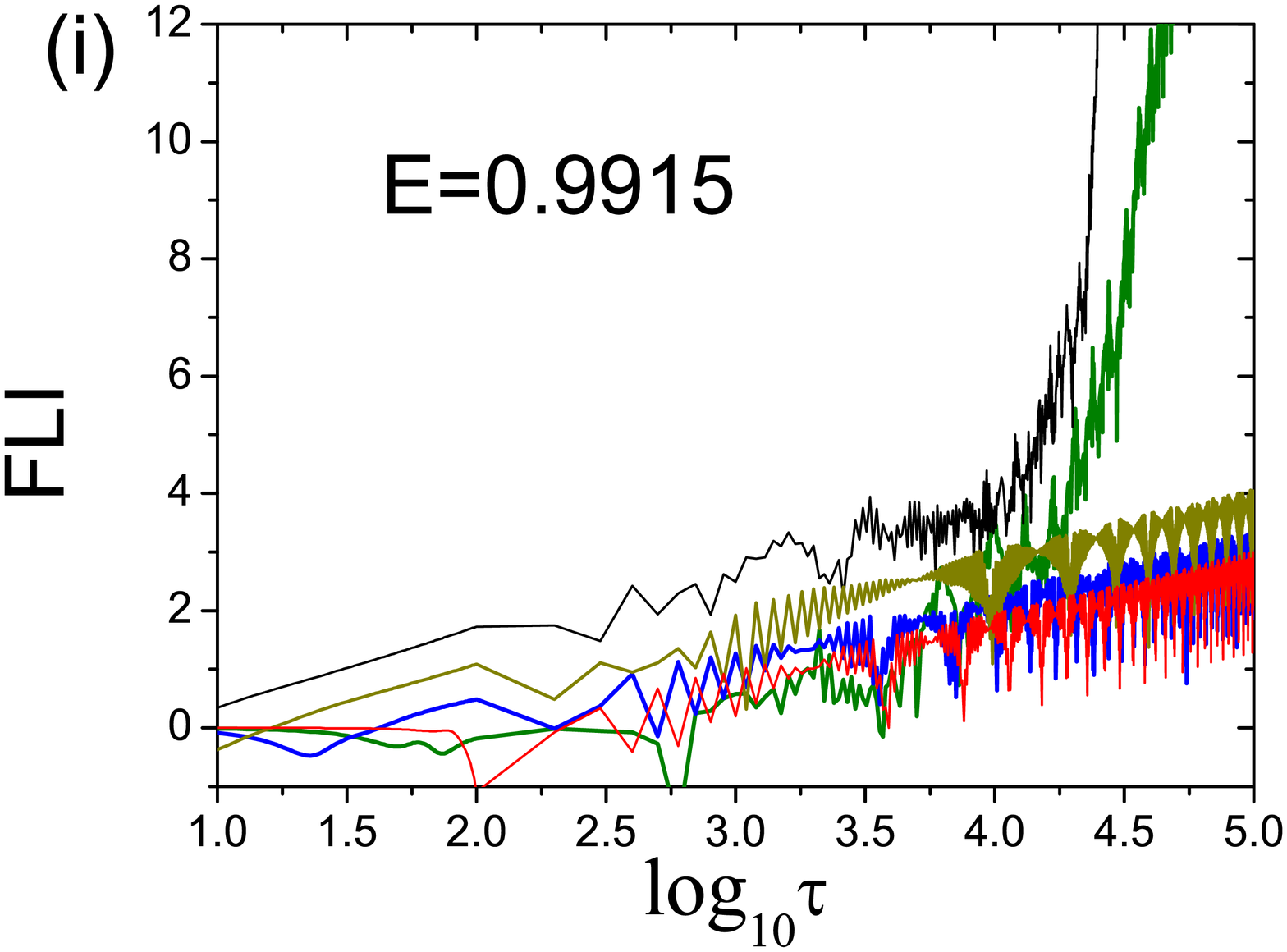}
\caption{(color online) (a)-(c): Poincar\'{e} sections at the plane $\theta=\pi/2$ with $p_{\theta}>0$ for three different values
of the energy $E$ in the system (22). The other parameters are $B=0.001$, $L=3.6$ and $q=0$, and the initial values $p_{r}=0$ and $\theta=\pi/2$ are fixed.
All orbits in panel (a) are regular KAM tori. There is a chaotic orbit with the initial value $r=10$ in panel (b). Two orbits
with the initial values $r=10$ and $r=55$ are chaotic in panel (c). (d)-(f): Lyapunov exponents $\lambda$ corresponding to the orbits in panels (a)-(c).
(g)-(i): Fast Lyapunov indictors (FLIs) corresponding to the orbits in panels (a)-(c).
}} \label{fig4}
\end{figure*}

\begin{figure*}
\center{
\includegraphics[scale=0.2]{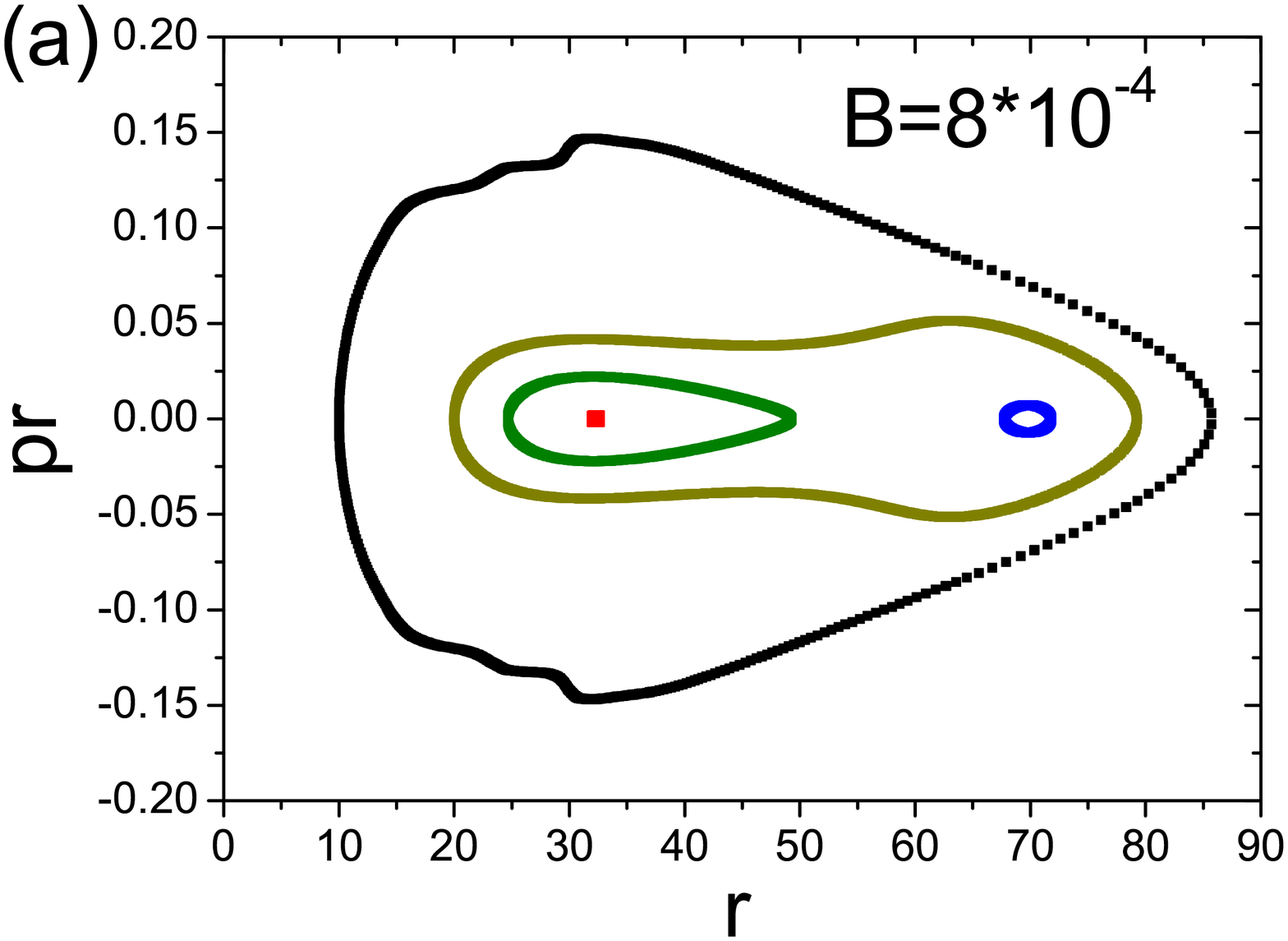}
\includegraphics[scale=0.2]{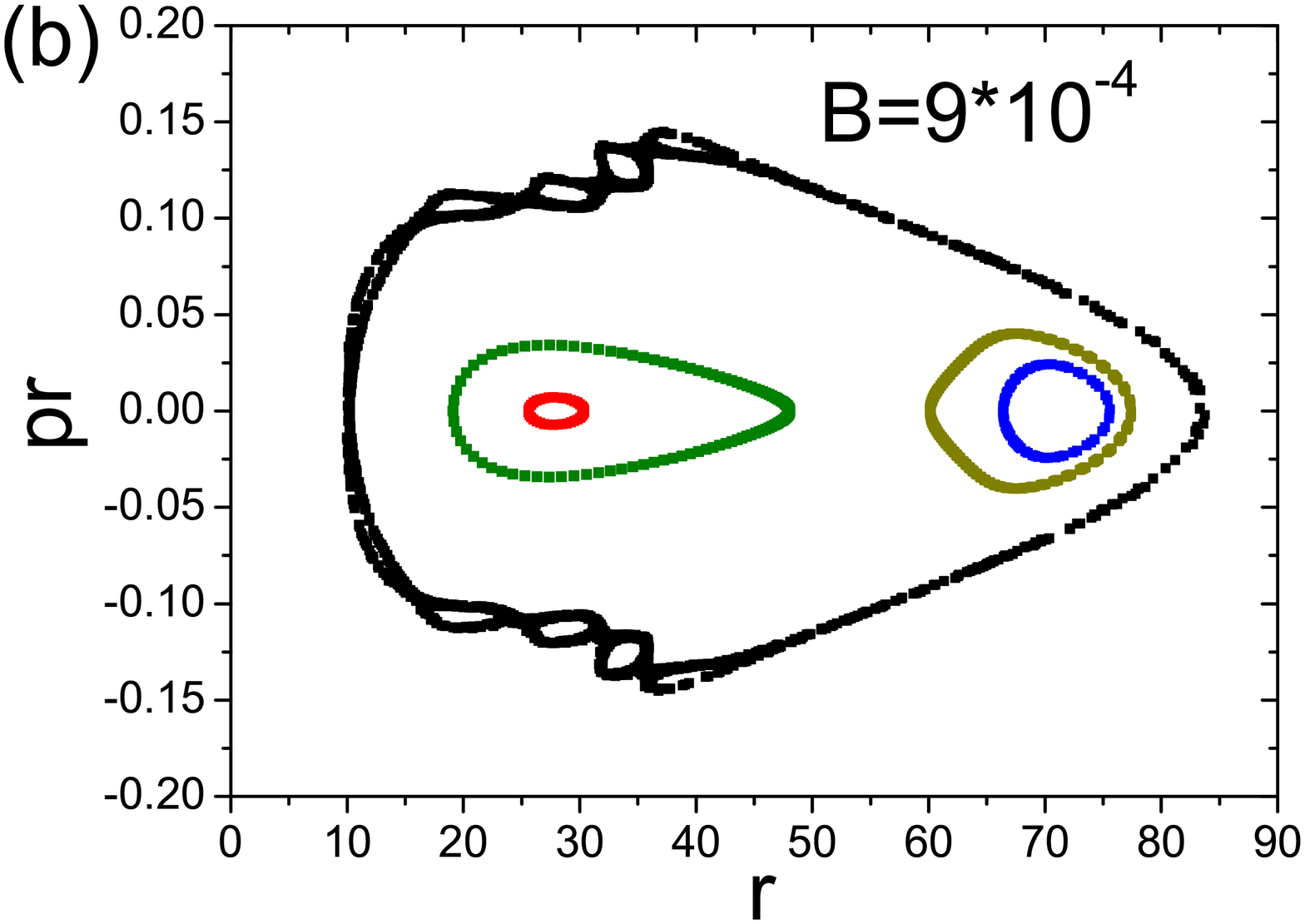}
\includegraphics[scale=0.2]{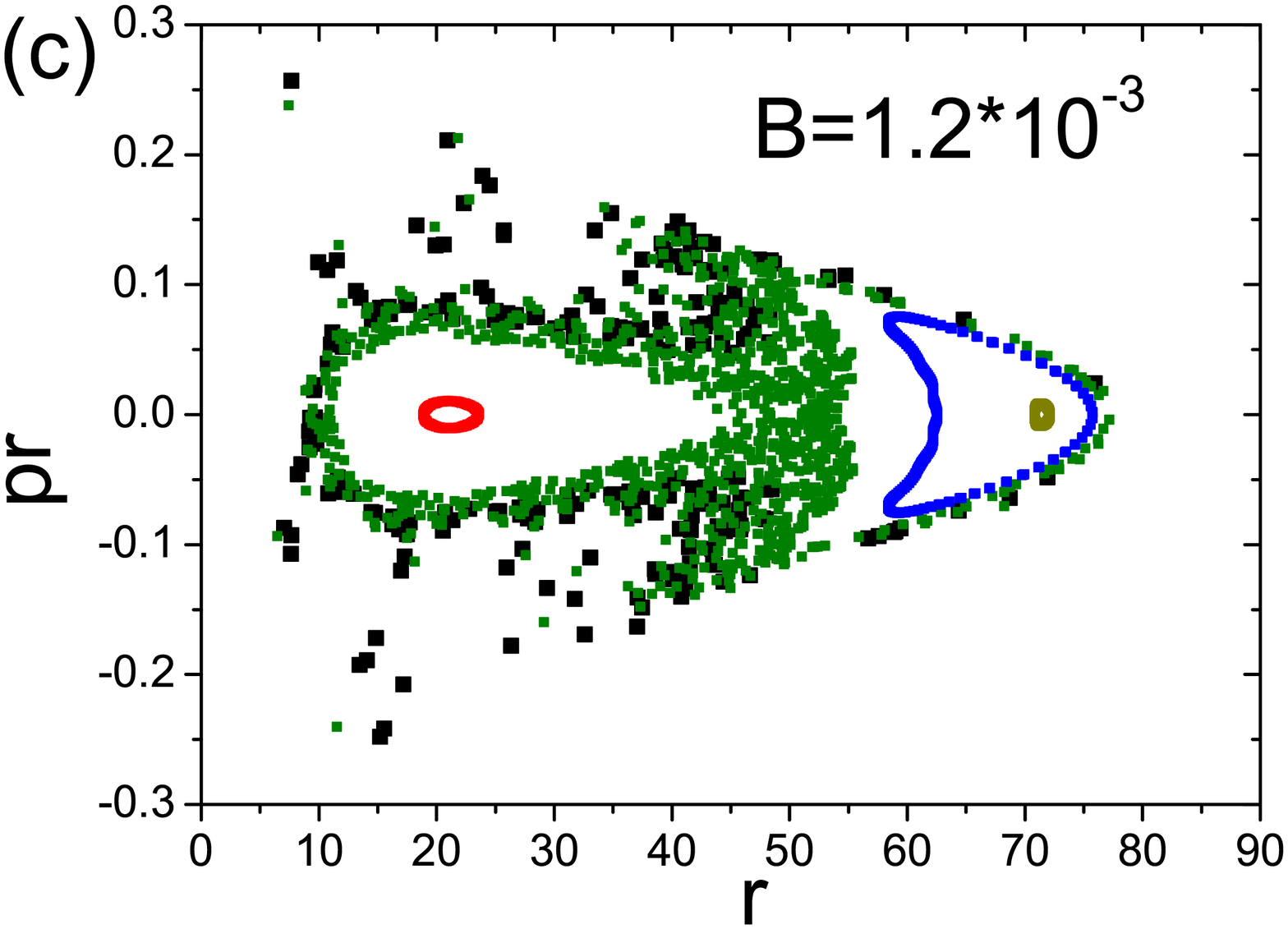}
\caption{(color online) Poincar\'{e} sections for three different values of the magnetic filed $B$. The other parameters are
$E=0.991$, $L=3.6$ and $q=0$.
}} \label{fig5}
\end{figure*}

\begin{figure*}
\center{
\includegraphics[scale=0.2]{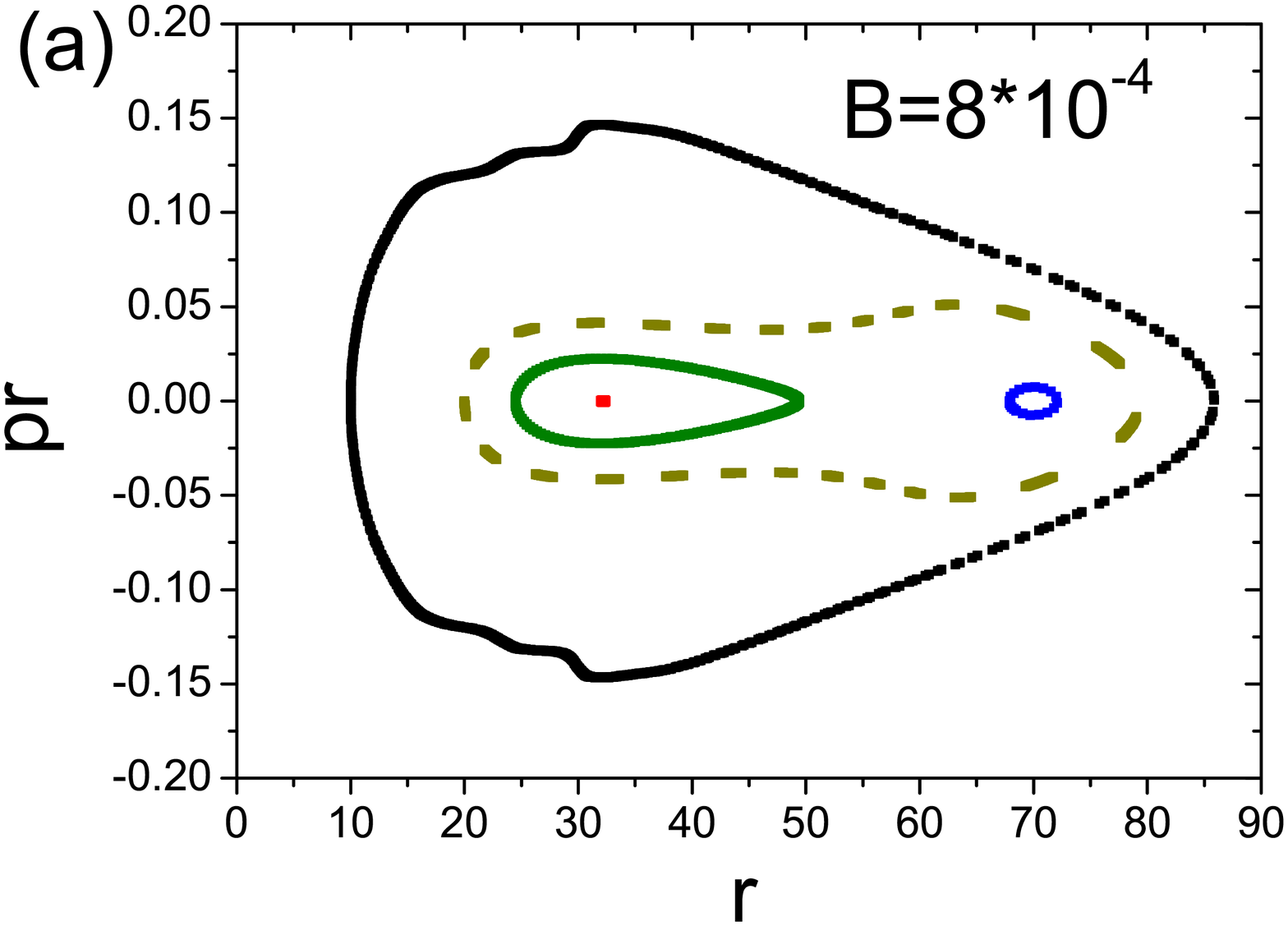}
\includegraphics[scale=0.2]{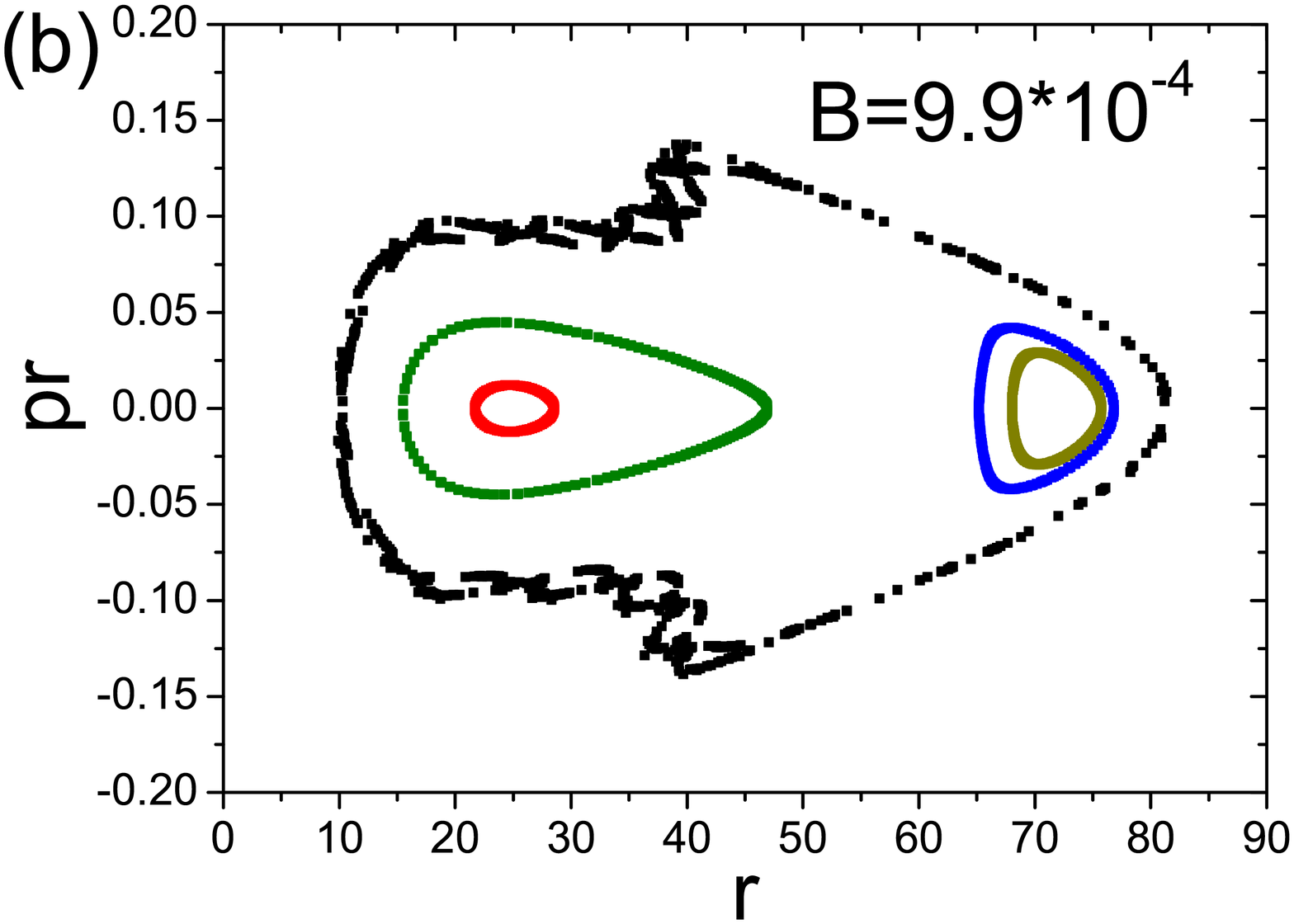}
\includegraphics[scale=0.2]{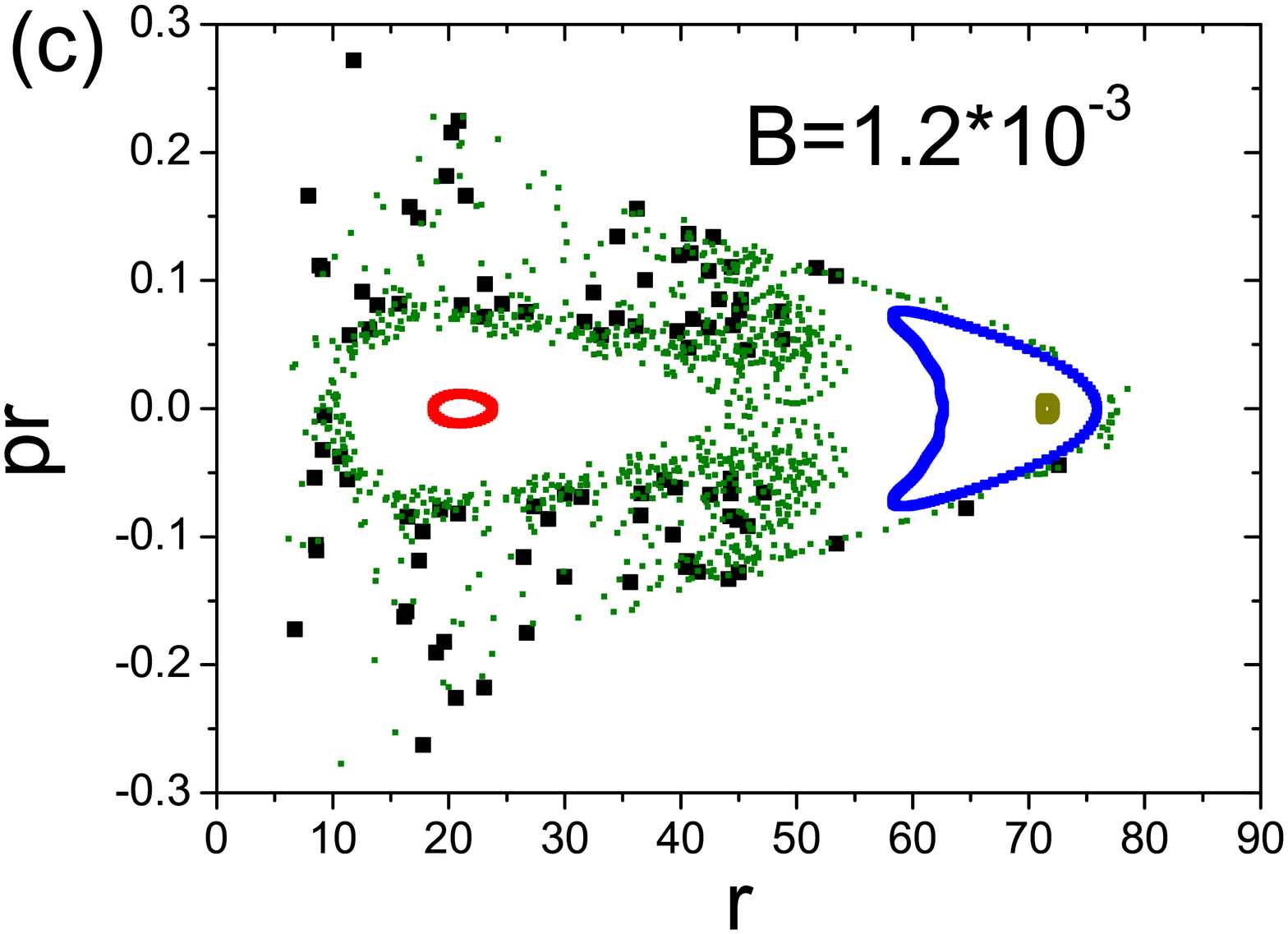}
\caption{(colour online) Same as Fig. 5 but the charge $q=0.01$.
}} \label{fig6}
\end{figure*}

\begin{figure*}
\center{
\includegraphics[scale=0.2]{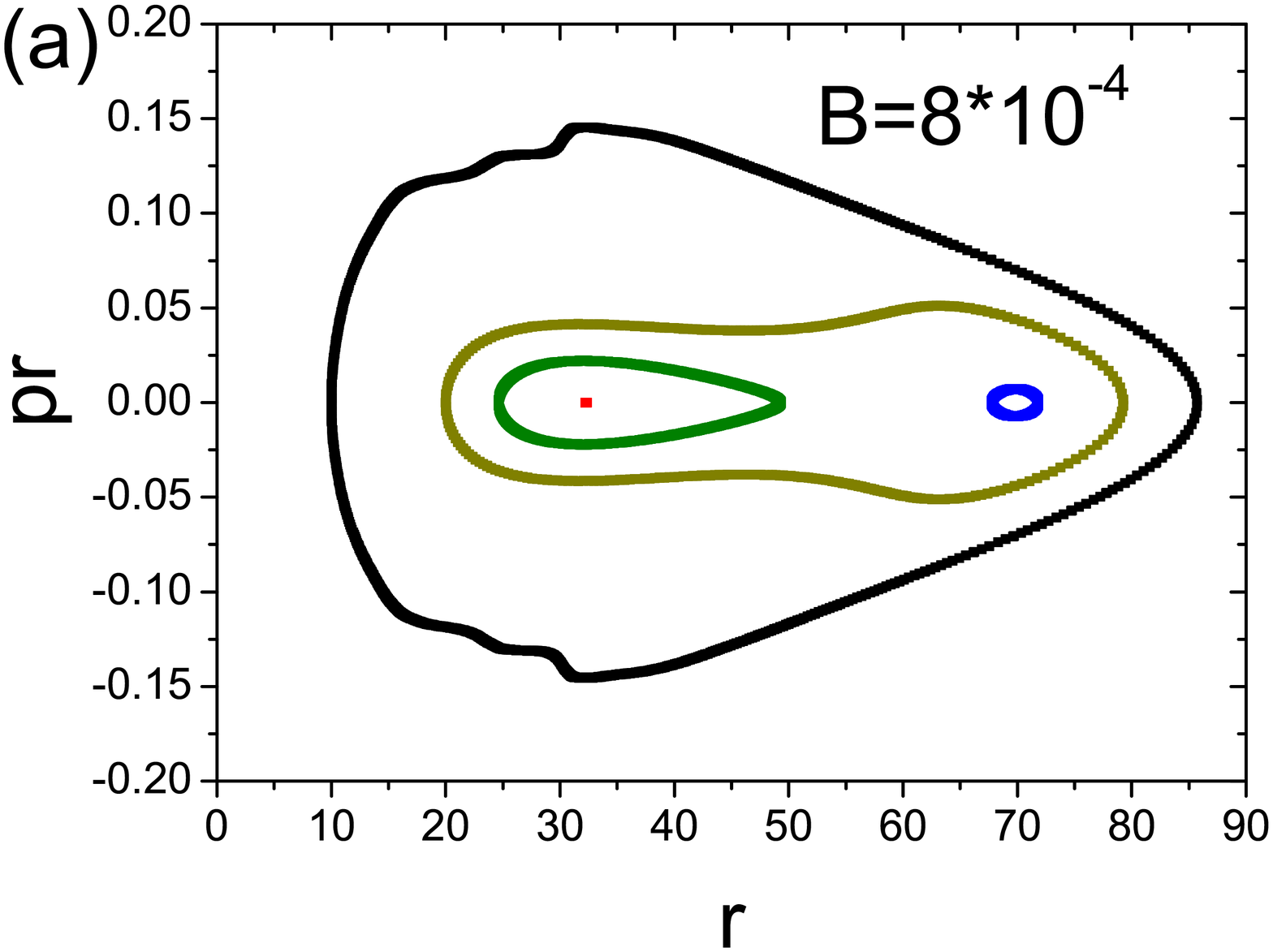}
\includegraphics[scale=0.2]{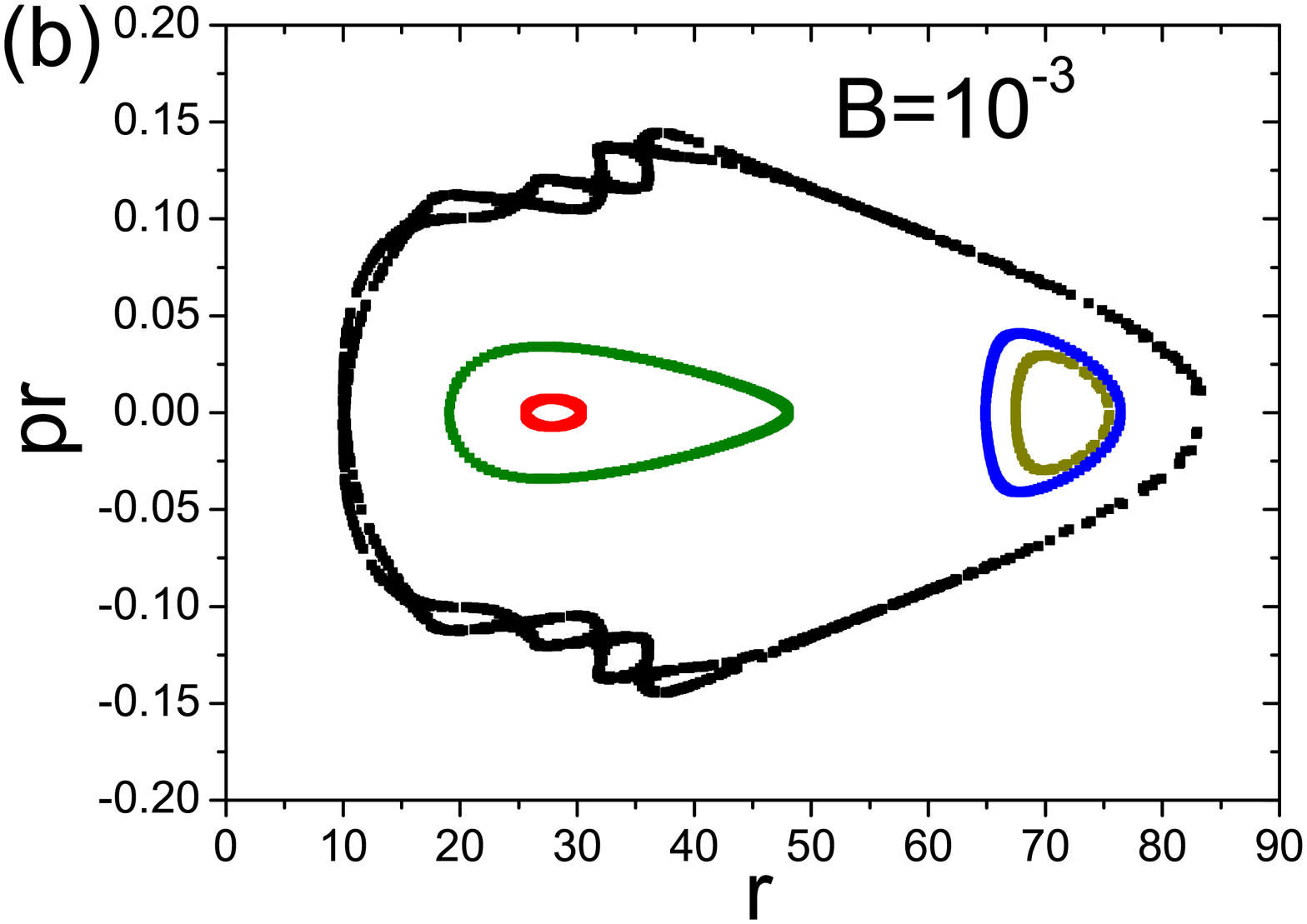}
\includegraphics[scale=0.2]{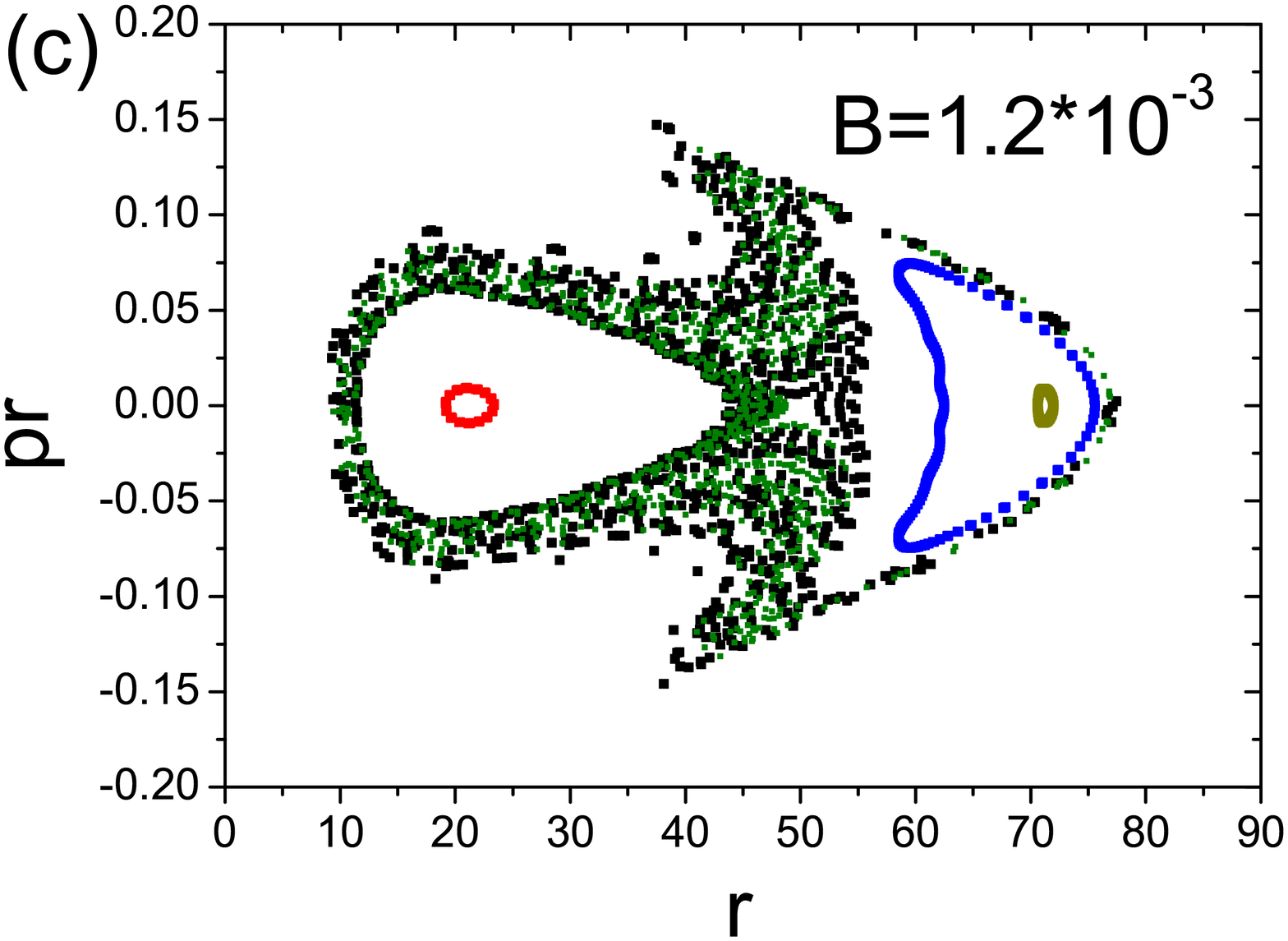}
\caption{(color online) Same as Fig. 6 but the charge $q=-0.01$.
}} \label{fig7}
\end{figure*}

\begin{figure*}
\center{
\includegraphics[scale=0.2]{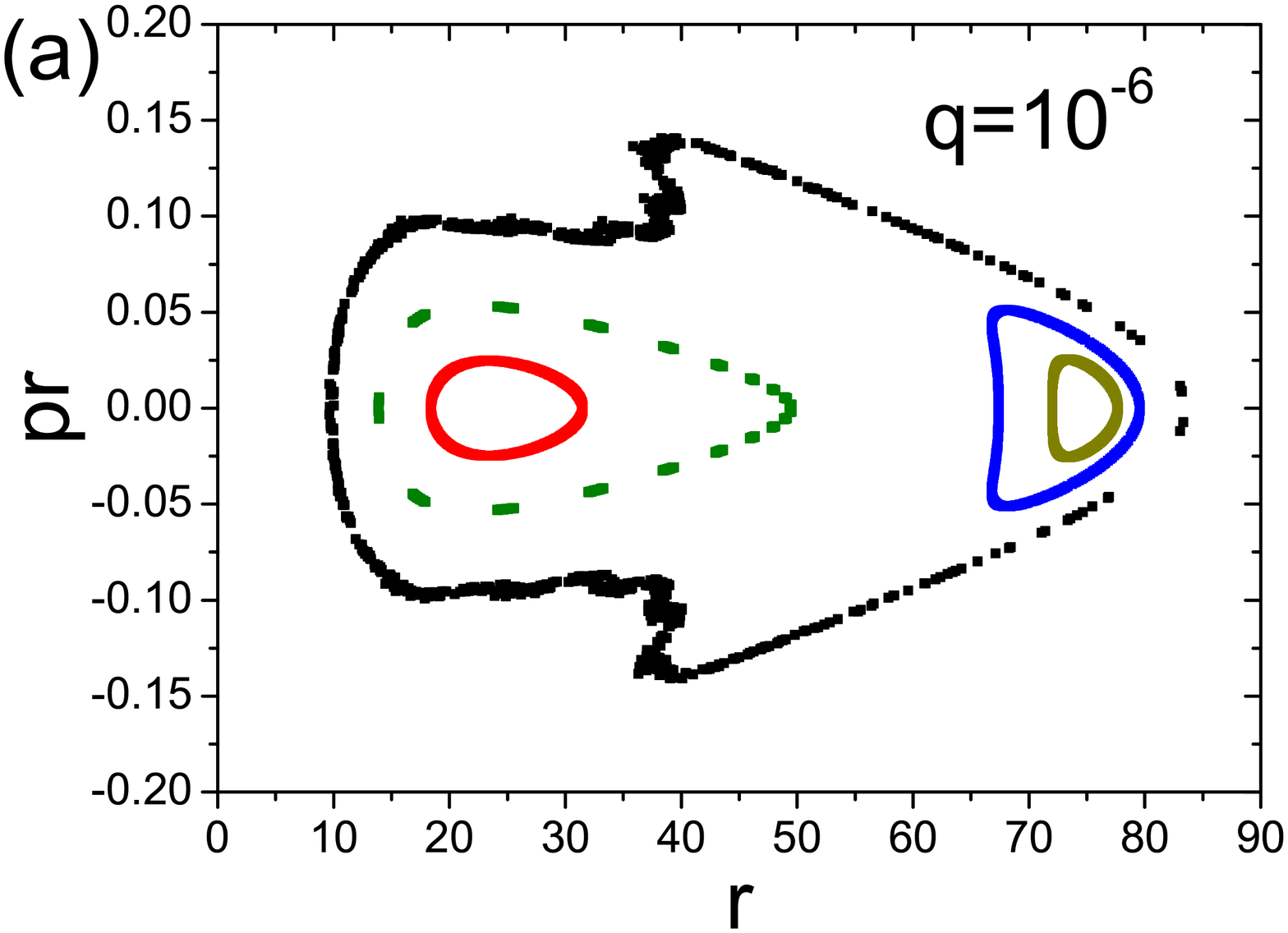}
\includegraphics[scale=0.2]{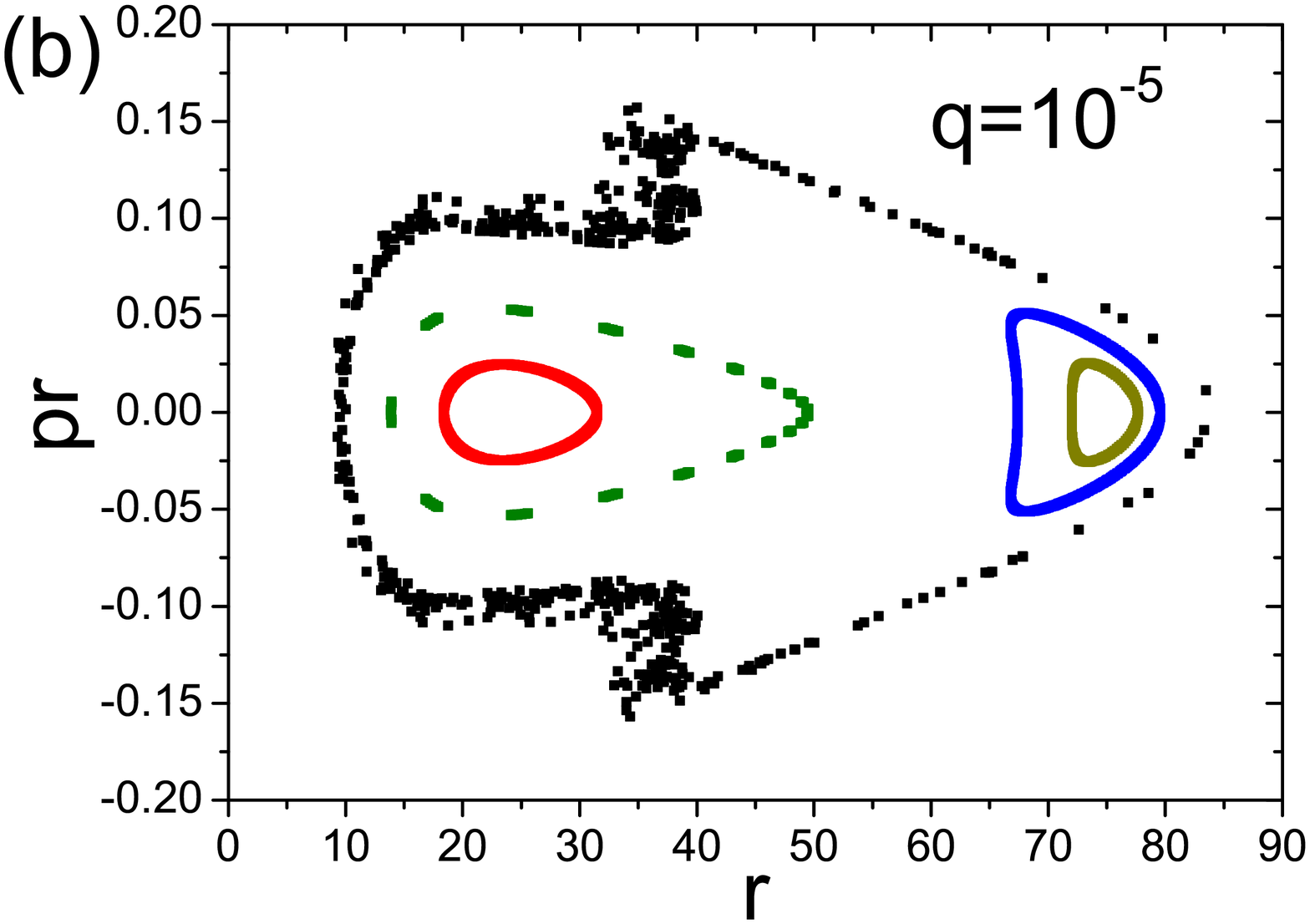}
\includegraphics[scale=0.2]{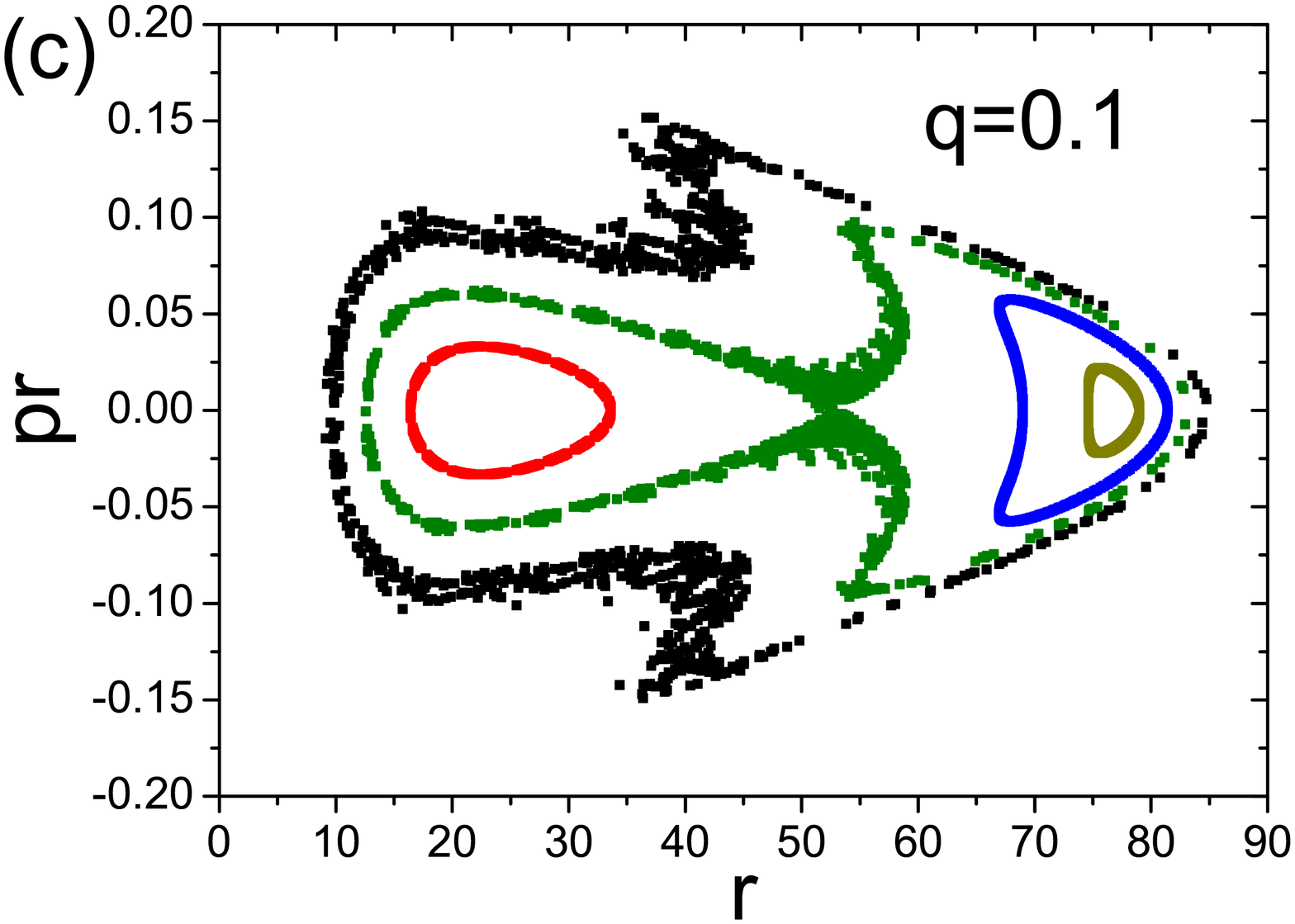}
\caption{(color online) Poincar\'{e} sections for three different values of the charge $q$. The other parameters are $E=0.9913$, $L=3.8$
 and $B=0.001$.
}} \label{fig8}
\end{figure*}

\begin{figure*}
\center{
\includegraphics[scale=0.2]{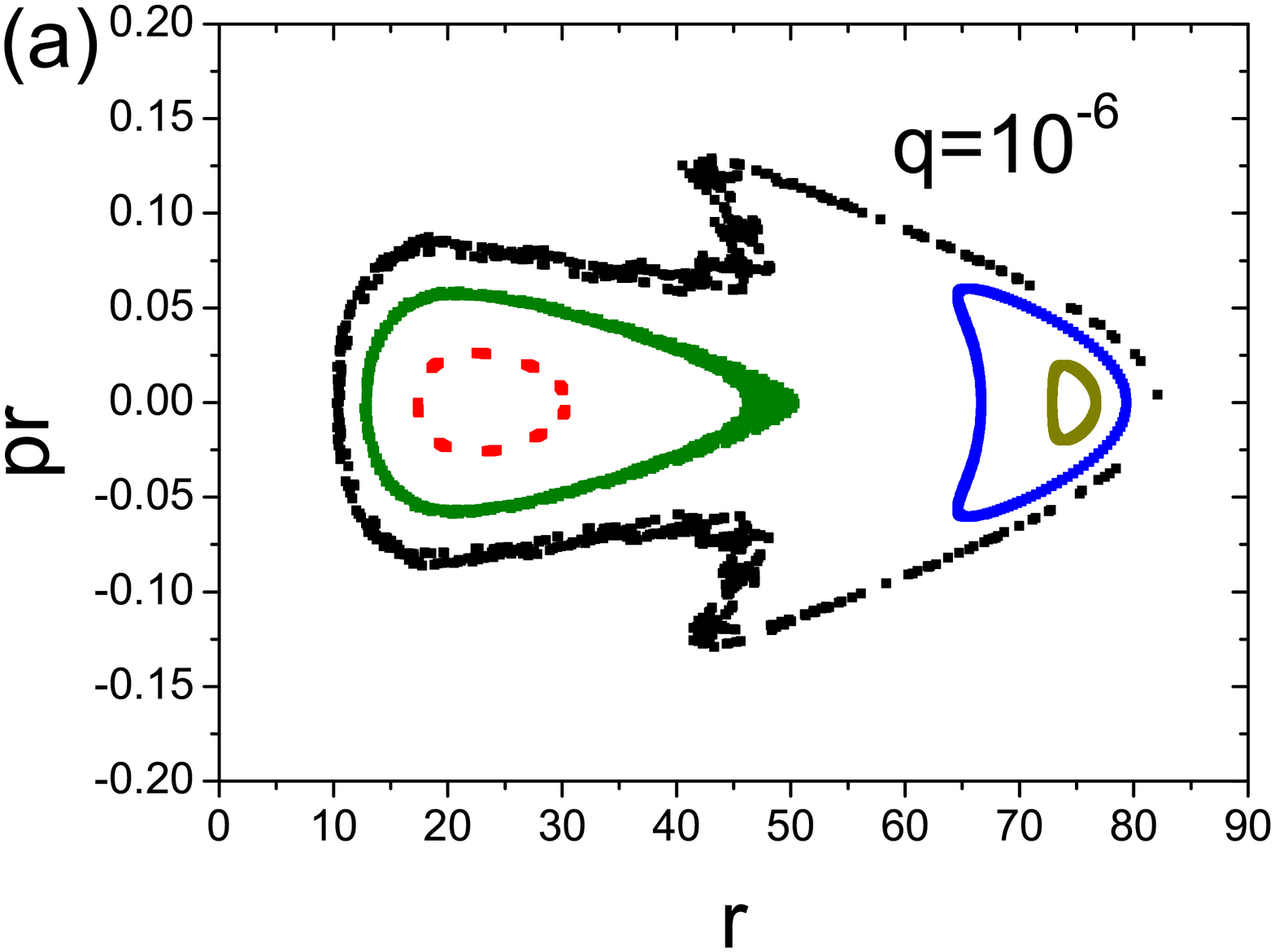}
\includegraphics[scale=0.2]{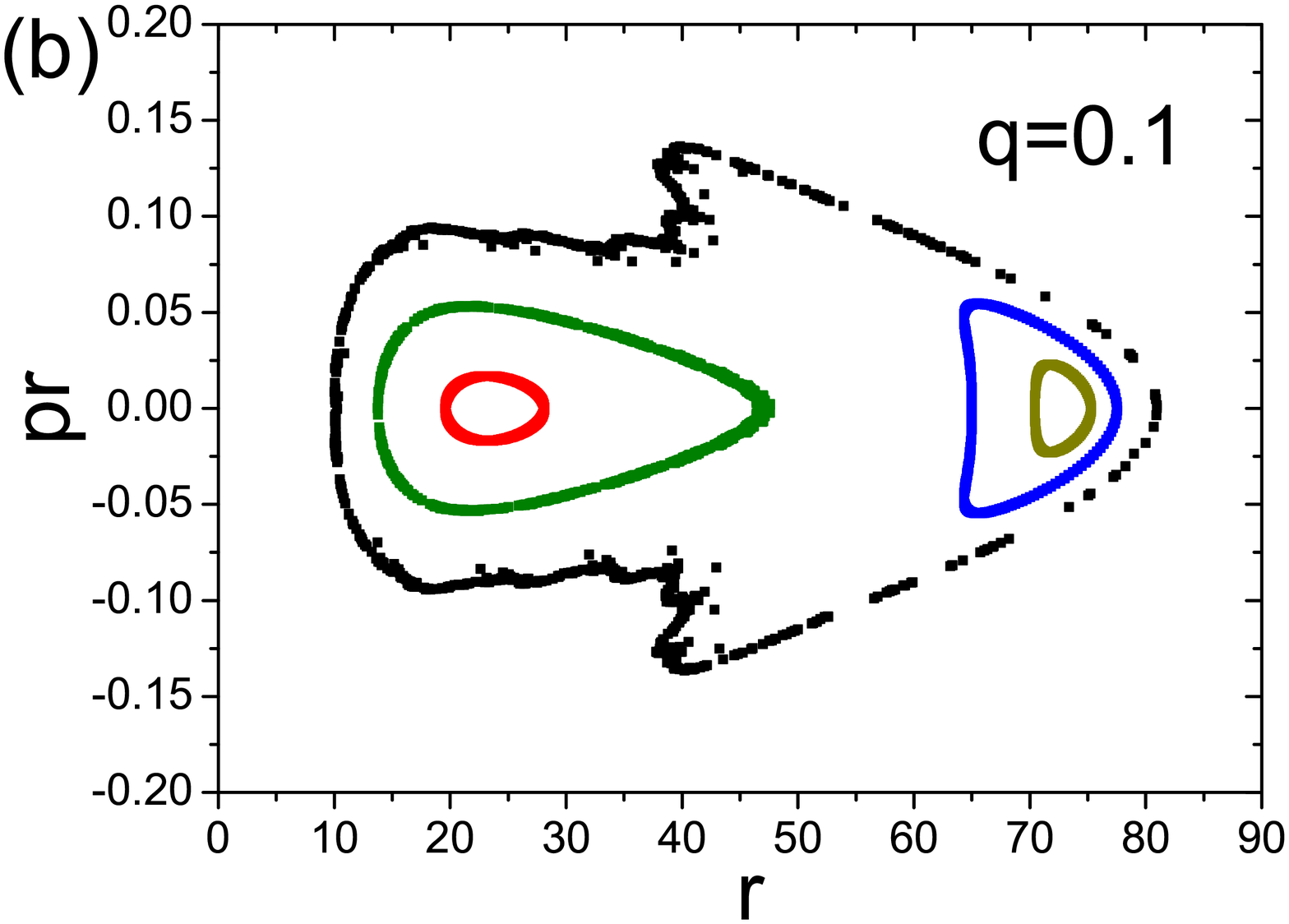}
\includegraphics[scale=0.2]{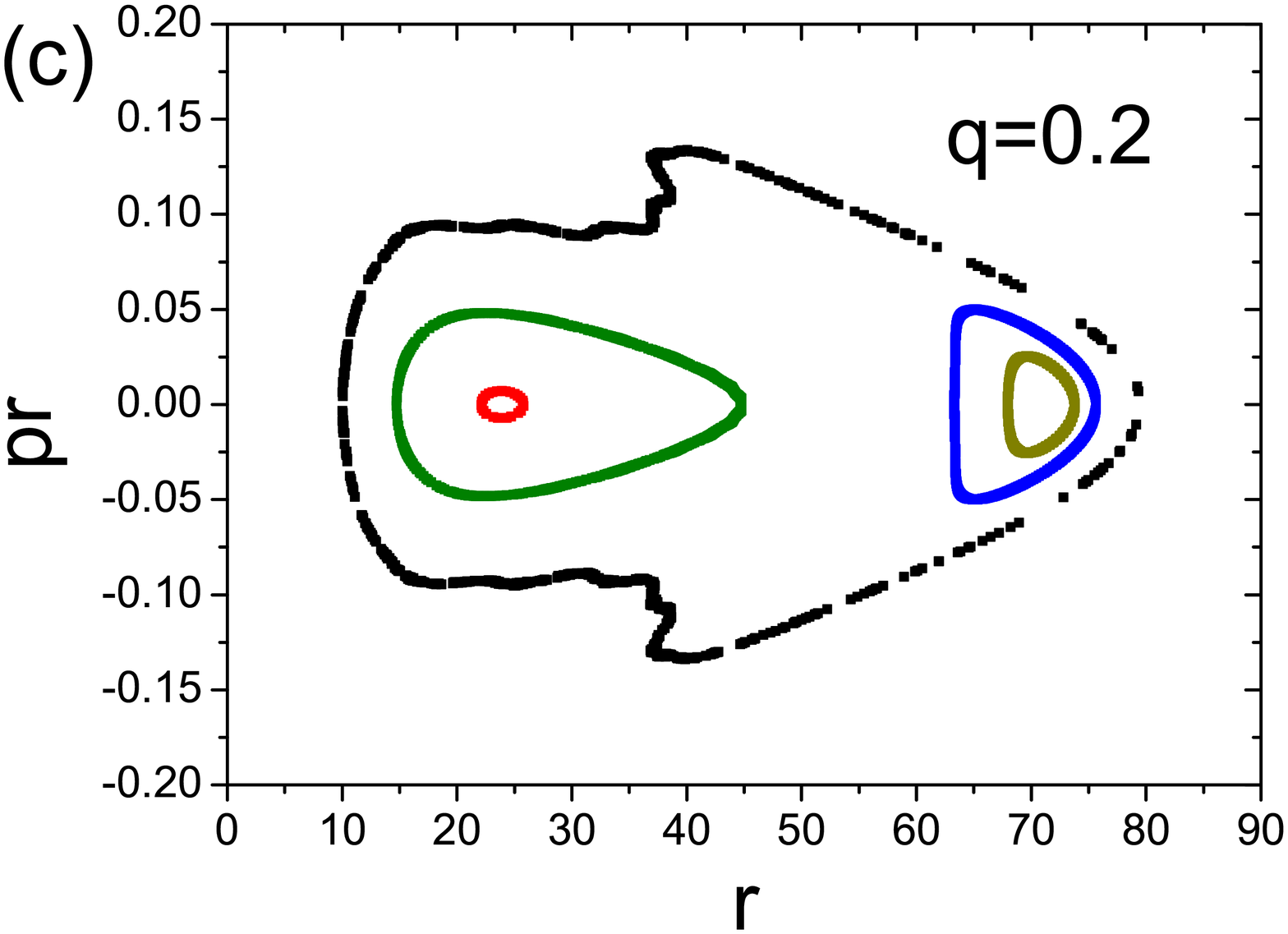}
\caption{(color online) Same as Fig. 8 but the negative magnetic filed $B=-0.001$ is given.
}} \label{fig9}
\end{figure*}

\end{document}